\documentclass[aip,graphicx]{revtex4-1}

\usepackage{amsmath,amssymb,xcolor}
\newcommand{\mat}[1]{\ensuremath\mathbf{#1}}
\renewcommand{\vec}[1]{\ensuremath\mathbf{#1}}
\newcommand{\matspace}[3]{\ensuremath\mathbb{#1}^{#2 \times #3}}
\newcommand{\nbas}[0]{\ensuremath N_b}
\newcommand{\ngrid}[0]{\ensuremath N_g}

\newcommand{\NCCL}[0]{\texttt{NCCL}~}

\usepackage{afterpage}

\newif\ifexporttikz
\newif\ifusetikz

\exporttikzfalse  
\usetikzfalse     

\ifexporttikz
    \usepackage{tikz}
    \usepackage{pgfplots}
    \pgfplotsset{compat=newest}
    \usepgfplotslibrary{external}
    \usepgfplotslibrary{fillbetween}
    \usetikzlibrary{patterns, intersections,arrows}
\else
    \ifusetikz
      \usepackage{tikzexternal}
      \tikzsetexternalprefix{fig-export/}
    \else
      \usepackage{graphicx}
    \fi
\fi

\ifusetikz
  
  \newcommand{\importlocalfigure}[1]{%
    \figname{#1}
    \input{fig-src/#1.tex}
  }
\else
  \newcommand{\importlocalfigure}[1]{%
    \includegraphics{fig-export/#1.pdf}
  }
\fi

\usepackage{newfloat,algcompatible}
\usepackage{algpseudocode}
\DeclareFloatingEnvironment[
    fileext=loa,
    listname=List of Algorithms,
    name=Algorithm,
    placement=tbhp,
]{algorithm}

\usepackage[capitalize]{cleveref}
\crefname{figure}{Fig.}{Figs.}
\Crefname{figure}{Figure}{Figures}
\crefname{table}{Tab.}{Tabs.}
\Crefname{table}{Table}{Tables}
\crefname{equation}{Eq.}{Eqs.}
\Crefname{equation}{Equation}{Equations}
\crefname{section}{Sec.}{Secs.}
\Crefname{section}{Section}{Sections}
\crefname{paragraph}{Sec.}{Secs.}
\Crefname{paragraph}{Section}{Sections}
\crefname{algorithm}{Alg.}{Algs.}
\Crefname{algorithm}{Algorithm}{Algorithms}
\crefname{theorem}{Thm.}{Thms.}
\Crefname{theorem}{Theorem}{Theorems}
\crefname{corollary}{Cor.}{Cors.}
\Crefname{corollary}{Corollary}{Corollaries}

\draft 

\begin{document}


\title{Distributed Memory, GPU Accelerated Fock Construction for
Hybrid, Gaussian Basis Density Functional Theory}



\author{David B. Williams-Young}
\email[]{dbwy@lbl.gov}
\affiliation{Applied Mathematics and Computational Research Division, Lawrence Berkeley National Laboratory, Berkeley, CA 94720}

\author{Andrey Asadchev}
\affiliation{Department of Chemistry, Virginia Tech, Blacksburg, VA 24061}

\author{Doru Thom Popovici}
\affiliation{Applied Mathematics and Computational Research Division, Lawrence Berkeley National Laboratory, Berkeley, CA 94720}

\author{David Clark}
\affiliation{NVIDIA Corporation, Santa Clara, CA 95051}

\author{Jonathan Waldrop}
\affiliation{Chemical and Biological Sciences Division, Ames National Laboratory, Ames, IA 50011}

\author{Theresa Windus}
\affiliation{Chemical and Biological Sciences Division, Ames National Laboratory, Ames, IA 50011}
\affiliation{Department of Chemistry, Iowa State University, Ames, IA 50011}

\author{Edward F. Valeev}
\affiliation{Department of Chemistry, Virginia Tech, Blacksburg, VA 24061}

\author{Wibe A. de Jong}
\affiliation{Applied Mathematics and Computational Research Division, Lawrence Berkeley National Laboratory, Berkeley, CA 94720}


\date{\today}

\begin{abstract}
With the growing reliance of modern supercomputers on
accelerator-based architectures such a GPUs, the development and optimization
of electronic structure methods to exploit these massively parallel resources
has become a recent priority.
While significant strides have been made in the development of GPU accereated, 
distributed memory algorithms for many-body (e.g. coupled-cluster)
and spectral single-body (e.g. planewave, real-space and finite-element density functional theory [DFT]),
the vast majority of GPU-accelerated Gaussian atomic orbital methods have focused on shared memory 
systems with only a handful of examples pursuing massive parallelism on distributed memory
GPU architectures. In the present work, we present a set of distributed memory algorithms for the
evaluation of the Coulomb and exact-exchange matrices for hybrid Kohn-Sham DFT with
Gaussian basis sets via direct density-fitted (DF-J-Engine) and seminumerical (sn-K) methods, respectively. The absolute performance and strong scalability of the developed methods 
are demonstrated on systems ranging from a few hundred to over one thousand atoms using up to 128 NVIDIA
A100 GPUs on the Perlmutter supercomputer.
\end{abstract}


\maketitle 

\section{Introduction}
\label{sec:intro}

Since its inception, quantum chemistry has relied on its ability to quickly
adapt to an ever evolving landscape of computer architectures to enable the
next generation of scientific applications. A particular
emphasis has been placed on the leverage of distributed memory parallelism to enable \emph{ab initio}
simulations of large molecular systems on the world's largest supercomputers
\cite{janssen2008parallel,deJong2010utilizing,calvin21_many,gavini2022roadmap}.
The past two decades have been no different, with an enormous research effort
having been afforded to targeting the latest introduction into the
high-performance computing (HPC) ecosystem: graphics processing units
(GPU)\cite{gordon20_novel,gordon20_modern,gavini2022roadmap}. The use of GPUs in quantum
chemistry is relatively long standing, with applications ranging from
single-body methods such as Hartree-Fock (HF) and density functional (DFT)
theories \cite{%
yasuda08_accelerating,
kalinowski17_arbitrary,
kussmann17_employing,laqua20_highly,kussmann15_preselective,
kussmann17_hybrid,kussmann21_highly,laqua22_accelerating,
ufimtsev09_quantum2,luehr16_gaussian,
williams20_on,williams21_achieving,ahmed21_performance,
asadchev11_new,barca20_high,barca20_scaling,barca21_faster,barca20_high,
manathunga21_harnessing,manathunga20_parallel,manathunga2023quantum,
maintz2011speeding,wang2011large,hacene2012accelerating,
giannozzi2020quantum,apra20_nwchem,wilkinson13_porting,
andrade13_real,hakala13_parallel,
das2022dft,
genovese09_density,
van16_gpu,
yoshikawa19_gpu,
huhn20_gpu
}, 
to many-body methods\cite{calvin21_many} such as 
coupled-cluster theory \cite{%
deprince2011coupled,deprince16_iterative,
asadchev2013fast,
kaliman17_new,
fales2020performance,hohenstein21_gpu,
ma2010acceleration,ma16_perturbative,
peng19_coupled,
pototschnig21_implementation
}, 
many-body perturbation theory \cite{%
olivares10_accelerating,olivares16_gpu,
song16_atomic,
maurer14_a,
bykov17_the,
barca20_q,barca2021enabling
}, 
and configuration interaction and complete active space theories
\cite{%
hohenstein2015atomic,fales2020efficient,
mullinax2019heterogeneous,
straatsma2020gronor,
maris2022accelerating
} 
to name a few. 
Despite significant advances in the development of
GPU-accelerated quantum chemistry software, these efforts are not yet as mature
as their central processing unit (CPU) counterparts, and growing
requirements for the desired scale and accuracy of \emph{ab initio} simulations
require further development in this area.  As such, the pursuance of improved,
GPU-accelerated quantum chemistry methods capable of leveraging the latest
advances in modern HPC remains an active area of research.

With the growing
reliance of modern exascale HPC systems on accelerator architectures,\cite{Kothe19exascale,alexander20_exascale} recent years have seen the
deployment of GPUs in distributed memory systems. This deployment has in turn been accompanied by its own 
unique set of optimization challenges. In particular, the increased local processing rate of
GPU-accelerated compute nodes has exposed bottlenecks involving communication and
load imbalance which were less apparent on the massively parallel CPU systems
of years past \cite{ashby2010opportunities,amarasinghe2011exascale,alexander20_exascale}.
As necessitated by the predominant availability of consumer-grade,
gaming-oriented GPU hardware, early development of GPU-accelerated quantum
chemistry methods focused on shared memory systems where host and device memory
spaces, while disjoint, are accessible to one another without the need for
communication across distributed memory networks. 
Due to their relative computational cost, many-body methods (e.g.,
coupled cluster theory \cite{ma2010acceleration,ma16_perturbative,peng19_coupled,pototschnig21_implementation})
and spectral single-body methods (e.g., 
planewave \cite{maintz2011speeding,wang2011large,hacene2012accelerating,giannozzi2020quantum,apra20_nwchem,wilkinson13_porting}, 
real-space \cite{andrade13_real,hakala13_parallel},
finite-element \cite{das2022dft},
and wavelet \cite{genovese09_density} discretizations of DFT) were among the
first to be successfully ported to distributed memory GPU architectures. The
success of these developments has been, in large part, enabled by the prevalent
use of GPU optimized libraries for common operations,  such as 
(multi-)linear algebra \cite{calvin2015scalable,calvin2015task,lyakh2015efficient,kim2019code,kaliman17_new} and 
fast Fourier transforms (FFT)~\cite{popovici2021systematic, ayala2020heffte, franchetti2018fftx}, in these applications. On the other hand, atomic orbital (AO) single-body methods, such as Gaussian, Slater, and numerical atomic orbital HF and DFT, present a unique
challenge for GPU optimization due to their significant dependence
on domain-specific kernels (e.g. AO integrals and highly-specialized numerical integration techniques) which exhibit less-regular compute patterns than those
common operations aforementioned. 
For Gaussian basis DFT in particular, kernels involving
manipulation of the AO electron repulsion integral (ERI) tensor 
to form the Coulomb and exchange matrices are among the most
cumbersome for GPU architectures. As such, the vast majority of both 
shared\cite{yasuda08_accelerating,%
luehr2011dynamic,luehr16_gaussian,
kalinowski17_arbitrary,
laqua20_highly,
kussmann21_highly,
asadchev10_696,asadchev11_new,barca20_high,
miao13_acceleration,miao15_acceleration,manathunga20_parallel
}
and distributed\cite{%
kussmann15_preselective,kussmann17_hybrid,
manathunga21_harnessing,
johnson22_multinode
}
memory Gaussian DFT GPU efforts have centered around optimization of these 
terms.

The primary innovations of the present work concern the evaluation of Coulomb and exact-exchange  matrices. Although  the naive (based on 2-body 4-center ERI) approach for evaluation of these contributions is simple and thus used by majority of implementations on distributed  heterogeneous platforms,\cite{kussmann15_preselective,kussmann17_hybrid,barca21_faster} their steep asymptotic scaling complicates their application to large molecular systems. For the Coulomb matrix, the use of 4-center integrals results in suboptimal $\mathcal{O}(N^2)$ cost, whereas fast approaches with $\mathcal{O}(N)$ for the Coulomb potential evaluation are well known\cite{VRG:greengard:1987:JCP,VRG:white:1994:CPL}. More importantly, even if combined with the fast $\mathcal{O}(N)$ treatment of the long-range contributions to the potential, the use of 4-center integrals for the near-field contributions is still prohibitively expensive. The solution is well known also: clever optimizations (like early integral digestion\cite{VRG:rezaahmadi:1995:CPL,VRG:white:1996:JCP,VRG:adams:1997:JCP,VRG:shao:2000:CPL}) and/or numerical approximations like density fitting\cite{VRG:whitten:1973:JCP,VRG:vahtras:1993:CPL} can dramatically reduce the cost of Coulomb matrix evaluation such that even with naive $\mathcal{O}(N^2)$ evaluation its cost is dwarfed by the cost of the exact exchange.

While with proper screening the exact exchange evaluation is $\mathcal{O}(N)$ in system size, the use of 4-center integrals, again, results in suboptimally-high prefactor. Thus, there has been a recent resurgence of 
interest in the development of numerical methods for the evaluation of exact exchange 
for molecular systems. In these methods, one of the two ERI coordinate integrations  is replaced
by a numerical integration. The general concept for this approach has been around for
over 30 years, beginning with the pseudospectral method on Friesner 
and co-workers
\cite{friesner1985solution,ringnalda1990pseudospectral} in the early 1980s.
This class of techniques was revisited in the early 2000s as the 
chain-of-spheres exchange (COSX) method  of Neese, \emph{et al}, \cite{neese2009efficient}
and more recently as the seminumerical exchange (sn-K) method of Laqua, \emph{et al}.\cite{laqua20_highly}
The primary differences between these approaches are in their consideration of
numerical sparsity, spatial locality, and in their attempts to reduce
the required size of the numerical quadrature to attain a particular accuracy.
We will adopt the sn-K nomenclature in the following. The sn-K algorithm is
particularly attractive for GPU hardware as there is a natural expression
of vectorization in the numerical quadrature which is superior to that of 
analytical integral methods. Significant progress has been made in the development of GPU sn-K algorithms for 
shared memory systems including multiple GPUs per node.\cite{laqua20_highly,laqua21_accelerating,kussmann21_highly,laqua22_accelerating} 
These methods have demonstrated a sizable performance improvement over analytical integral methods on
GPU architectures, but have yet to be explored in the context of distributed memory parallelism.

In addition to the Coulomb and exact exchange terms, hybrid AO DFT
methods also require the evaluation of the exchange-correlation (XC) potential matrix
by numerical integration methods due to the non-linear nature of the XC energy functional. 
In contrast to AO ERI methods, numerical integration methods developed for molecular
DFT are much simpler to port to GPU architectures \cite{yasuda08_accelerating,williams20_on} and 
are able to heavily utilize optimized BLAS libraries to attain near peak floating point performance
on modern GPU hardware \cite{williams21_achieving}. Recently, there have been a number of works 
which have addressed key optimization challenges pertaining to the distributed memory evaluation
of the XC potential on GPU accelerated computing clusters \cite{williams20_on,manathunga21_harnessing}.
In this work, we extend the  parallel integration infrastructure developed for the XC potential by
the authors\cite{williams20_on} to treat the 
Gaussian basis sn-K method
on distributed memory architectures. 

The remainder of this work is organized as follows.
In \cref{sec:theory} we review the salient aspects 
of Gaussian basis DFT necessary to develop parallel
algorithms for the evaluation of the Coulomb (\cref{sec:rij}) and exact-exchange (\cref{sec:snK}) matrices. In \cref{sec:batch_screen,sec:load_balance},
we described the extension of the distributed memory
integration procedure developed for the XC potential (\cref{sec:xc}) to treat sn-K on GPU clusters. 
In \cref{sec:results}, we demonstrate the strong
scaling performance of the developed methods for a
range of systems and, finally, discuss future outlook
for further development of distributed memory GPU
algorithms based on the methods presented in this work in \cref{sec:conclusions}.

\section{Theory and Implementation}
\label{sec:theory}
\subsection{The Hybrid Kohn-Sham Fock Matrix}
\label{sec:hybrid_dft}
Given a set of basis functions, 
$\mathcal{B} = \{\phi_\mu : \mathbb{R}^3\rightarrow\mathbb{R}\}_{\mu=1}^{\nbas}$, 
the hybrid Kohn-Sham (KS) Fock matrix , $\mat{F}\in\matspace{R}{\nbas}{\nbas}$, is given by
\cite{szabo2012modern,yang95_dftbook}
\begin{equation}
\mat{F} = \mat{h} + \mat{J}[\mat{D}] - c_x \mat{K}[\mat{D}] + \mat{V}^{xc}[\mat{D}],
\end{equation}
where $\mat{h}$ is the core Hamiltonian which contains the free-particle (kinetic)
Hamiltonian and the electron-nuclear interaction potential, and $\mat{D}$ is the
basis-discretized one-particle density matrix. $\mat{J}$, $\mat{K}$, and $\mat{V}^{xc}$
are the density-dependent (as denoted by $\left[\,\cdot\,\right]$) Coulomb, exact-exchange,
and exchange-correlation (XC) potential matrices, respectively, which describe the mean-field 
electron-electron interactions in the hybrid KS model as modulated
by the exact-exchange parameter $c_x\in\mathbb{R}^+$. For real-valued basis functions and density matrices,
these matrices take the forms \cite{yang95_dftbook}
\begin{align}
    J_{\mu\nu} &= \sum_{\lambda\kappa} (\mu\nu\vert\lambda\kappa) D_{\lambda\kappa} 
      \label{eq:full_j} \\
    K_{\mu\nu} &= \sum_{\lambda\kappa} (\mu\lambda\vert\nu\kappa) D_{\lambda\kappa} 
      \label{eq:full_k}\\
    V^{xc}_{\mu\nu} &= \int_{\mathbb{R}^3}\mathrm{d}^3\vec{r}\,
      \phi_\mu(\vec{r}) \frac{\delta E^{xc}[\rho(\vec{r})]}{\delta \rho(\vec{r})} \phi_\nu(\vec{r})
      \label{eq:exact_vxc}
\end{align}
where $E^{xc}$ is the XC energy functional evaluated at the electon density, 
$\rho:\mathbb{R}^3\rightarrow\mathbb{R}$
\begin{equation}
\rho(\vec{r}) = \sum_{\mu\nu} D_{\mu\nu}\phi_\mu(\vec{r})\phi_\nu(\vec{r})
\end{equation}
and
\begin{equation}
(\mu\lambda\vert\nu\kappa) = \iint_{\mathbb{R}^3}\mathrm{d}^3\vec{r}\mathrm{d}^3\vec{r}'\, 
  \frac{\phi_\mu(\vec{r})\phi_\lambda(\vec{r})\phi_\nu(\vec{r}')\phi_\kappa(\vec{r}')}{\vert\vec{r} - \vec{r}'\vert}
\end{equation}
is the electron-repulsion integral (ERI) tensor.

In this work, we take $\mathcal{B}$ to be comprised of contracted, atom-centered 
Gaussian-type orbitals (GTO), however, we note that the general principles presented here may be 
extended to other atom-centered ans\"atze, such as Slater-type orbitals (STO), as 
well. We denote the atomic center of $\phi_\mu$ as $\mathbf{R}_\mu$ in the following. GTO Fock matrix construction is dominated by the three density-dependent 
terms which must be evaluated for each new density in e.g., an SCF optimization or
dynamics simulation. In the remainder of this section, we examine the
scalable evaluation of these terms on distributed memory GPU architectures. 

\subsection{The Coulomb Matrix via Density-Fitting J-Engine}
\label{sec:rij}

Efficient evaluation of the Coulomb potential \cref{eq:full_j} takes advantage of the  density fitting (DF; also known as the Resolution-of-the-Identity) approximation of the two-electron integrals:
\begin{align}
   &(\mu\nu\vert\lambda\kappa) \approx \sum_{KL} (\mu\nu\vert K) (\mat{V}^{-1})_{KL} (L \vert \lambda\kappa)
\end{align}
where 3- and 2-center Coulomb integrals involving the (auxiliary) density-fitting basis {$\{\chi_K:\mathbb{R}^3\rightarrow\mathbb{R}\}_{K=1}^{N_\mathrm{aux}}$} were introduced:
\begin{align}
(\mu\nu\vert K) &= \iint_{\mathbb{R}^3}\mathrm{d}^3\vec{r}\mathrm{d}^3\vec{r}'\, 
\frac{\phi_\mu(\vec{r})\phi_\lambda(\vec{r})\chi_K(\vec{r}')}{\vert\vec{r} - \vec{r}'\vert} \\
V_{LK} &= 
\iint_{\mathbb{R}^3}\mathrm{d}^3\vec{r}\mathrm{d}^3\vec{r}'\, 
  \frac{\chi_L(\vec{r})\chi_K(\vec{r}')}{\vert\vec{r} - \vec{r}'\vert}.
\end{align}
DF approximation of \cref{eq:full_j} leads to the following factorization of the Coulomb potential:
\begin{align}
V_L &= \sum_{\lambda\kappa}(L \vert \lambda\kappa) D_{\lambda\kappa}, \label{eq:V_df} \\
D_K &= \sum_L (\mat{V}^{-1})_{KL} V_L, \label{eq:P_df} \\
J_{\mu\nu}  &\overset{\mathrm{DF}}{\approx} \sum_K (\mu\nu\vert K) D_K, \label{eq:J_df}
\end{align}
with $V_L$ and $D_K$ representing the Coulomb potential of the exact density $\rho$ and the {\em robust} approximation of $\rho$ in the DF basis, respectively. DF-based evaluation of $J$ costs $\mathcal{O}(N^3)$ in the traditional approach where the Cholesky factorization of the positive-definite matrix $\mathbf{V}$ is computed and stored (this allows to amortize its cost over the SCF iterations). 
In practice, however, the cost is largely controlled by the evaluation of $\mathcal{O}(N^2)$ nonnegligible Coulomb 3-center AO integrals in \cref{eq:V_df,eq:J_df}. While for 3-center AO integrals dedicated optimization of AO integral evaluation is possible,\cite{VRG:ahlrichs:2004:PCCPP,VRG:hollman:2015:JCP,VRG:valeev:2020:JCP} including specific developments
for the GPU architectures,\cite{doi:10.1021/acs.jctc.2c00995} optimal evaluation of \cref{eq:V_df,eq:J_df} involves blurring the line between the integral evaluation and density contraction via several related ideas\cite{VRG:rezaahmadi:1995:CPL,VRG:white:1996:JCP,VRG:adams:1997:JCP,VRG:shao:2000:CPL} that is often dubbed the {\em J-engine} approach. The original J-engine approaches were demonstrated in the context of \cref{eq:full_j}; Kussmann et al.\cite{kussmann21_highly} recently illustrated the utility of the McMurchie-Davidson-based\cite{VRG:mcmurchie:1978:JCP} J-engine\cite{VRG:shao:2000:CPL} in the DF context. Here we only briefly recap the DF-based J-engine (``DF-J-engine'') formalism using the established notation\cite{obara1986efficient} for Gaussian AO integrals.

An uncontracted primitive Cartesian Gaussian with exponent $\zeta_a \in\mathbb{R}^+$ and non-negative integer Cartesian ``quanta'' $\mathbf{a} \equiv \{a_x, a_y, a_z \}$ centered at $\mathbf{R}_a \equiv \{A_x, A_y, A_z\}$ will be denoted by
\begin{align}
    \phi_\mathbf{a} ({\bf r}) \equiv x_A^{a_x} y_A^{a_y} z_A^{a_z} \exp(-\zeta_a r_A^2),
\end{align}
with $\mathbf{r}_A \equiv \{x_A, y_A, z_A\}, x_A \equiv x - A_x$, etc.
$l_\mathbf{a} \equiv a_x + a_y + a_z \geq 0$ is colloquially referred to as the ``angular momentum'' of a Gaussian. Closely related to the Cartesian Gaussian is a Hermite Gaussian:
\begin{align}
    \Lambda_\mathbf{\tilde{a}} ({\bf r}) \equiv \left(\frac{\partial}{\partial x_A}\right)^{\tilde{a}_x} \left(\frac{\partial}{\partial y_A}\right)^{\tilde{a}_y} \left(\frac{\partial}{\partial z_A}\right)^{\tilde{a}_z} \exp(-\zeta_a r_A^2).
\end{align}
A primitive Cartesian Gaussian and a product of two primitive Cartesian Gaussians can be expressed as linear combinations of Hermite Gaussians,
\begin{align}
\label{eq:cart2herm-1}
    \phi_\mathbf{a}({\bf r}) = & \sum_{\tilde{a}_x=0}^{\tilde{a}_x \leq a_x} E_{a_x}^{\tilde{a}_x}  \sum_{\tilde{a}_y=0}^{\tilde{a}_y \leq a_y} E_{a_y}^{\tilde{a}_y} \sum_{\tilde{a}_z=0}^{\tilde{a}_z \leq a_z} E_{a_z}^{\tilde{a}_z} \Lambda_\mathbf{\tilde{a}}(\mathbf{r}) \equiv \sum_\mathbf{\tilde{a}} E_\mathbf{a}^\mathbf{\tilde{a}} \Lambda_\mathbf{\tilde{a}}(\mathbf{r}), \\
\label{eq:cart2herm-2}
    \phi_\mathbf{a}({\bf r})\phi_\mathbf{b}({\bf r}) = & \sum_{\tilde{p}_x=0}^{\tilde{p}_x \leq a_x+b_x} \left(E_x\right)_{a_x b_x}^{\tilde{p}_x}  \sum_{\tilde{p}_y=0}^{\tilde{p}_y \leq a_y + b_y} \left(E_y\right)_{a_y b_y}^{\tilde{p}_y} \sum_{\tilde{p}_z=0}^{\tilde{p}_z \leq a_z + b_z} \left(E_z\right)_{a_z b_z}^{\tilde{p}_z} \Lambda_\mathbf{\tilde{p}}(\mathbf{r}) \equiv \sum_\mathbf{\tilde{p}} E_{\mathbf{a} \mathbf{b}}^\mathbf{\tilde{p}} \Lambda_\mathbf{\tilde{p}}(\mathbf{r}) ,
\end{align}
with $\zeta_p \equiv \zeta_a + \zeta_b$, $\mathbf{R}_p \equiv \frac{\zeta_a \mathbf{R}_a + \zeta_b \mathbf{R}_b}{\zeta_a + \zeta_b}$, and  $E_\mathbf{a}^\mathbf{\tilde{a}} \equiv \prod_{i={x,y,z}} E_{a_i}^{\tilde{a}_i} $, $E_\mathbf{a b}^\mathbf{\tilde{p}} \equiv \prod_{i={x,y,z}}\left(E_i\right)_{a_i b_i}^{\tilde{p}_i} $. The Hermite-to-Cartesian transformation coefficients are evaluated straightforwardly by recursion.\cite{VRG:mcmurchie:1978:JCP}

The use of \cref{eq:cart2herm-1,eq:cart2herm-2} allows to express a 3-center Coulomb integral over primitive Cartesian Gaussians as a linear combination,
\begin{align}
\label{eq:abc_to_hermite}
(\mathbf{a} \mathbf{b}|\mathbf{c}) = & \sum_{\mathbf{\tilde{p}}, \mathbf{\tilde{c}}} E_\mathbf{a b}^\mathbf{\tilde{p}} E_\mathbf{c}^\mathbf{\tilde{c}} (\mathbf{\tilde{p}} | \mathbf{\tilde{c}}),
\end{align}
of the Coulomb integral between two primitive Hermite Gaussians:
\begin{align}
(\mathbf{\tilde{p}} | \mathbf{\tilde{c}}) \equiv & \, 
\iint_{\mathbb{R}^3}\mathrm{d}^3\vec{r}\mathrm{d}^3\vec{r}'\, 
\frac{\Lambda_\mathbf{\tilde{p}}(\vec{r}) \Lambda_\mathbf{\tilde{c}}(\vec{r}')}{\vert\vec{r} - \vec{r}'\vert} .
\end{align}
The latter can be evaluated directly,
\begin{align}
\label{eq:pc}
(\mathbf{\tilde{p}} | \mathbf{\tilde{c}}) \equiv (-1)^{l_\mathbf{\tilde{c}}} (\mathbf{\tilde{p}}+\mathbf{\tilde{c}})^{(0)},
\end{align}
from the auxiliary integral,
\begin{align}
(\mathbf{\tilde{r}})^{(m)} \equiv \left(\frac{\partial}{\partial x_R}\right)^{\tilde{r}_x} \left(\frac{\partial}{\partial y_R}\right)^{\tilde{r}_y} \left(\frac{\partial}{\partial z_R}\right)^{\tilde{r}_z} (\mathbf{0})^{(m)},
\end{align}
with $(\mathbf{0})^{(m)}$ related to the Boys function $F_m(x)$:
\begin{align}
(\mathbf{0})^{(m)} \equiv \, & (-2 \rho)^m \frac{2 \pi^{5/2}}{\zeta_p \zeta_c \sqrt{\zeta_p+\zeta_c}} F_m(\rho |\mathbf{R}_p-\mathbf{R}_c|^2) , \\
F_m(x) \equiv \, & \int_0^1 \, \mathrm{d}y \, y^{2m} \exp(-x y^2), \label{eq:boys} \\
\rho \equiv \, & \frac{\zeta_p \zeta_c}{ \zeta_p + \zeta_c }.
\end{align}
Auxiliary integrals are evaluated recursively,
\begin{align}
\label{eq:md1}
    (\mathbf{\tilde{r}}+\mathbf{1}_i)^{(m)} = & \tilde{r}_i (\mathbf{\tilde{r}}-\mathbf{1}_i)^{(m+1)} + (P_i-C_i) (\mathbf{\tilde{r}})^{(m+1)},
\end{align}
starting from $(\mathbf{0})^{(m)}$.

Efficient evaluation of Eqs. \eqref{eq:V_df} and \eqref{eq:J_df} involves contracting densities with Hermite-to-Cartesian transformation coefficients of the ket functions (Eqs. \eqref{eq:cart2herm-2} and \eqref{eq:cart2herm-1}, respectively) to produce ``Hermite'' densities which can be stored and
reused for every bra function in \cref{eq:V_df,eq:J_df}, as illustrated here for a single primitive shell contribution to \cref{eq:V_df}:
\begin{align}
\sum_{\mathbf{a b}}
(\mathbf{c}|\mathbf{a} \mathbf{b}) D_{\mathbf{a b}} \overset{\mathrm{Eq.\,\eqref{eq:abc_to_hermite}}}{=} & \sum_\mathbf{a b}  \left( \sum_\mathbf{\tilde{c} \tilde{p}} E^\mathbf{\tilde{c}}_\mathbf{c} (\mathbf{\tilde{c}}|\mathbf{\tilde{p}})  E^\mathbf{\tilde{p}}_{\mathbf{a b}}  \right) D_{\mathbf{a b}} 
\label{eq:cabP-int}
\\
= & \sum_\mathbf{\tilde{c}} E^\mathbf{\tilde{c}}_\mathbf{c} 
\left( \sum_\mathbf{\tilde{p}} (\mathbf{\tilde{c}}|\mathbf{\tilde{p}}) D_\mathbf{\tilde{p}} \right) , \label{eq:cabP-jengine}
\end{align}
where
\begin{align}
\label{eq:Dp}
D_\mathbf{\tilde{p}} \equiv \sum_\mathbf{a b} E^\mathbf{\tilde{p}}_{\mathbf{a b}} D_{\mathbf{a b}}
\end{align}
and the order of evaluation is indicated by the parentheses.
Refactorization of \cref{eq:cabP-int} via \cref{eq:cabP-jengine} is the key idea of J-engine: it leads to great FLOP reduction which is easily rationalized as follows: instead of multiplying ``matrix'' $(\mathbf{\tilde{c}}|\mathbf{\tilde{p}})$ by ``matrix'' $E^\mathbf{\tilde{p}}_{\mathbf{a b}}$, then evaluating the inner product with ``vector'' $D_{\mathbf{a b}}$, in the J-engine factorization the inner product of ``matrix'' $(\mathbf{\tilde{c}}|\mathbf{\tilde{p}})$ with ``vector'' $D_\mathbf{\tilde{p}}$ is evaluated directly.
This can also be viewed as early ``digestion'' of the integrals; more general framework for early digestion of the integrals beyond the J matrix evaluation has been considered by Gill et al.\cite{VRG:adams:1997:JCP}

\Cref{eq:V_df,eq:J_df} of the DF-J-engine approach were implemented in the open-source \texttt{LibintX} library\cite{doi:10.1021/acs.jctc.2c00995,libintx} 
 and \cref{eq:P_df} was implemented using the TiledArray library for distributed tensor contractions.\cite{calvin2015scalable,tiledarray}
Evaluation of \cref{eq:V_df} in \texttt{LibintX} proceeds as follows:
\begin{enumerate}
\item {\bf CPU}: construct a batch of $\mathbf{ab}$ shell pairs with same $l_\mathrm{ket} = l_\mathbf{a}+l_\mathbf{b}$ such that $D_\mathbf{a b}$ is local, copy to mapped memory buffers;
\item {\bf GPU}: launch a kernel to transform current batch of $D_\mathbf{a b}$ to $D_\mathbf{\tilde{p}}$ (\cref{eq:Dp});
\item {\bf GPU}: launch kernels to evaluate Hermite integrals (\cref{eq:pc}) and their contributions to the Coulomb potential (\cref{eq:cabP-jengine}) for all shells $\mathbf{c}$ (one kernel per $l_\mathbf{c}$);
\item repeat for the next $\mathbf{ab}$ batch.
\end{enumerate}
Note after launching the GPU kernels, the CPU starts to work on preparing the density and metadata for the next batch, thus the GPU and CPU work overlaps. Note that no significant metadata is kept persistent between SCF iterations to allow GPU memory use by other application stages.

Evaluation of \cref{eq:J_df} in \texttt{LibintX} is only slightly more complicated:
\begin{enumerate}
\item {\bf CPU}: construct a batch of $\mathbf{ab}$ shell pairs with same $l_\mathrm{bra} = l_\mathbf{a}+l_\mathbf{b}$ such that $J_\mathbf{ab}$ is local;
\item {\bf GPU}: launch a kernel to evaluate $E^\mathbf{\tilde{p}}_\mathbf{ab}$;
\item {\bf GPU}: launch kernels to evaluate Hermite density $D_\mathbf{\tilde{c}} \equiv \sum_\mathbf{c} E^\mathbf{\tilde{c}}_\mathbf{c} D_\mathbf{c}$, Hermite integrals (\cref{eq:pc}), and their contributions to the Coulomb potential in the Hermite basis,
\begin{align}
J_\mathbf{\tilde{p}} = \sum_\mathbf{\tilde{c}} (\mathbf{\tilde{p}}|\mathbf{\tilde{c}}) D_\mathbf{\tilde{c}},
\end{align}
(this is analogous to the first step in \cref{eq:cabP-jengine}, but adapted to the evaluation of \cref{eq:J_df})
 for all shells $\mathbf{c}$ (one kernel per $l_\mathbf{c}$);
\item {\bf GPU}: launch kernel to transform $J_\mathbf{\tilde{p}}$ to $J_\mathbf{ab}$ via
\begin{align}
J_\mathbf{ab} = & \sum_\mathbf{\tilde{p}} E^\mathbf{\tilde{p}}_\mathbf{ab} J_\mathbf{\tilde{p}}
\end{align}
\item {\bf CPU}: store $J_\mathbf{ab}$ shell-sets
\item repeat for the next $\mathbf{ab}$ batch;
\end{enumerate}
CPU and GPU work is again overlapped, with the CPU thread starting to work on steps 1-3 of the next batch while the work on the previous batch is being completed; step 4 is scheduled only after completion of the previous batch's step 5.

The distributed memory work distribution in \texttt{LibintX}'s DF-J-engine is statically determined by the distribution of the user-provided density matrix among ranks and by the expected distribution of the J matrix result. This design decision is motivated by the desire to simplify the API of the \texttt{LibintX} library and avoid mandating the density to be in any particular data structure or distribution manner; the user provides the density to the library in the form of a C++ lambda function (closure) that provides one block of the density matrix at a time (or fails to provide it if it is not located on the current node). Production computations in \cref{sec:results}  provided density and stored the resulting J matrix as a block-sparse distributed array (\texttt{DistArray}) object of the \texttt{TiledArray} framework.\cite{calvin2015task,tiledarray}
Tiling of the density matrix was determined by the k-means-based clustering of the atoms,\cite{VRG:lewis:2016:JCTC}
with tiles divided evenly among ranks (for $p$ distributed ranks, the first $n^2/p$ tile ordinals assigned to rank 0, the next $n^2/p$ tile ordinals assigned to rank 1, etc.; note than due to sparsity actual tile counts per rank are lower and end up being problem dependent). In evaluation of \cref{eq:V_df}, each rank evaluates all contributions that involve the tiles of the density matrix that are screened our and that reside on that rank; each rank produces contributions to all elements of vector $V_L$, thus these contributions are reduced across all ranks before evaluating \cref{eq:P_df}.
A similar lambda-based mechanism is used by \texttt{LibintX} to return the distributed J matrix to the user and distribute the work in evaluation of \cref{eq:J_df}.

In the interest of keeping the focus on the distributed-memory parallelization, the kernel-level implementation details of \texttt{LibintX}'s DF-J-engine will be kept to a minimum here. Each GPU thread evaluates the entire set of ``1-center'' auxiliary integrals (\cref{eq:md1}), converts them to ``2-center'' integrals via \cref{eq:pc} and ``digests'' them via \cref{eq:cabP-jengine} (or its analog for \cref{eq:J_df}) for a single primitive shell. This allows the algorithm to completely eliminate thread divergences since every thread in a thread block performs exactly the same computation. This approach is radically different from the work distribution  when computing integrals\cite{doi:10.1021/acs.jctc.2c00995} in which multiple threads cooperatively evaluated shell components; this is possible due to the greatly reduced memory requirements of the DF-J-engine compared to the AO integral engine. A key feature of the kernel implementation is the use of compile-time (templates and \texttt{constexpr}) C++ programming instead of custom code generation, similarly to how the 3-center integral evaluation (rather than the J-engine) was implemented earlier.\cite{doi:10.1021/acs.jctc.2c00995}

\subsection{The Exchange-Correlation Potential Matrix}
\label{sec:xc}

Unlike $\mat{J}$, which may be assembled directly from analytical
integrals, the integrals involved in $\mat{V}^{xc}$ (\cref{eq:exact_vxc}) must be evaluated 
numerically due to the nonlinear nature of $E^{xc}$ and its functional derivatives. As detailed
elsewhere \cite{pople1992kohn,yasuda08_accelerating,burow11_linear,petrone2018efficient}, for atom centered bases, $\mat{V}^{xc}$ may be efficiently evaluated via
\begin{align}
    V^{xc}_{\mu\nu} &\approx \sum_{i\in\mathcal{Q}} \Phi_{\mu i} Z_{\nu i} + Z_{\mu i} \Phi_{\nu i} \label{eq:vxc_num}\\
    \Phi_{\mu i} &= \phi_\mu(\vec{r}_i) \label{eq:collocation}
\end{align}
where $\boldsymbol{\Phi}$ is the collocation matrix and $\mathcal{Q} = \{
(w_i,\vec{r}_i) \}_{i=1}^{\ngrid}$ is a numerical quadrature consisting of nodes,
$\vec{r}_i\in\mathbb{R}^3$, and grid weights, $w_i\in\mathbb{R}$. For
molecular calculations, due to the irregular nature of the integrands in the
vicinity of nuclear charges, $\mathcal{Q}$ is typically taken as the union of
spherical quadratures ($\mathbb{R}\times S^2$) origined at each atomic center
coupled with a modified weight partitioning scheme to account for overlapping regions. We refer the reader to Refs.
\citenum{becke88_a,stratmann96_achieving,laqua18_an,mura96_improved,murray93_quadrature,treutler95_efficient,gill03_radial,chien06_sg0,gill93_a}
for more comprehensive discussions regarding the construction of generic
molecular quadratures.
In this work, atomic grids are constructed according to the Mura-Knowles (MK) scheme \cite{mura96_improved} with Lebedev-Laikov\cite{lebedev76_quadratures} angular grids, and
we use the molecular weight partitioning scheme of Stratmann, \emph{et al} \cite{stratmann96_achieving}.
In addition, angular grids are radially pruned 
according to the scheme of Treutler, \emph{et al} 
\cite{treutler95_efficient}.

The auxiliary matrix $\mat{Z}\in\mathbb{R}^{\nbas\times\ngrid}$ is 
method-dependent; the form of which depends on $E^{xc}$. Within the spin-restricted generalized-gradient approximation (GGA),
$\mat{Z}$ takes the form \cite{pople1992kohn,burow11_linear}
\begin{align}
    &Z_{\mu i} = 
     \frac{w_i}{2} \left(E^\rho_i \Phi_{\mu i} + 
  4 E^\gamma_i
    \left(\nabla\rho(\mathbf{r}_i)\cdot\nabla\Phi_{\mu i}\right)
    \right), \label{eq:zmat} \\
    &E^{\rho/\gamma}_i = 
      \left.
      \frac{\partial\varepsilon(\rho(\vec{r}),\gamma(\vec{r}))}{\partial \rho/\gamma}
      \right\vert_{\vec{r} = \vec{r}_i} \label{eq:exc_grid}\\
    &\rho(\vec{r}_i) = \sum_\mu F_{\mu i} \Phi_{\mu i},\qquad
    \nabla\rho(\vec{r}_i) = 2\sum_\mu F_{\mu i} \nabla \Phi_{\mu i}, \label{eq:den_grid}\\
    &F_{\mu i} = \sum_\nu D_{\mu\nu} \Phi_{\nu i} \label{eq:fmat},
\end{align}
where $\varepsilon:\mathbb{R}^2\rightarrow\mathbb{R}$ is the GGA energy 
density which depends on $\rho$ and $\gamma = \nabla\rho\cdot\nabla\rho$. 
Similar expressions have been derived for spin-generalized and meta-GGA XC functionals \cite{pople1992kohn,egidi17_two,petrone2018efficient,kussmann21_highly}.
This evaluation scheme for $\mat{V}^{xc}$ is particularly attractive for GPU architectures
due to the fact that the most compute-intensive operations can be implemented
using level-3 (\cref{eq:vxc_num} via SYR2K and \cref{eq:fmat} via GEMM)
and level-1 (\cref{eq:den_grid} via DOT 
\footnote{Evaluation of the density and its gradient actually proceeds as a batched-DOT as discussed in \cref{sec:batch_screen}}) BLAS operations \cite{yasuda08_accelerating,williams20_on,kussmann21_highly}. As such, these operations can leverage 
heavily optimized GPU BLAS libraries, leaving only a small number of
DFT-specific kernels (e.g. \cref{eq:collocation,eq:zmat,eq:exc_grid})
to be optimized by the chemistry-domain developer. We will examine 
specific details regarding the use of GPU BLAS libraries for this application 
in \cref{sec:load_balance}.

\subsection{The Seminumerical Exact-Exchange Matrix}
\label{sec:snK}

Although $\mathbf{K}$ can be evaluated directly from Coulomb integrals as in \cref{eq:full_k}, this is rarely the most efficient approach due to the steep rise of the computational cost of the 4-center integrals with the angular momenta. 
As detailed elsewhere\cite{friesner1985solution,ringnalda1990pseudospectral,neese2009efficient,laqua20_highly},
the ERI tensor may be factorized on a
quadrature grid via
\begin{align}
&(\mu\lambda\vert\nu\kappa) \approx \frac{1}{2}\sum_{i\in\mathcal{Q}} w_i A_{\mu\lambda i} \Phi_{\nu i} \Phi_{\kappa i} + (\mu\lambda\leftrightarrow\nu\kappa)\, , 
  \label{eq:sn_eri}\\
&A_{\mu\lambda i} = \int_{\mathbb{R}^3}\mathrm{d}^3\vec{r}\, \frac{\phi_\mu(\vec{r})\phi_\lambda(\vec{r})}{\vert \vec{r} - \vec{r}_i\vert}, \label{eq:3c_grid} 
\end{align}
where $\mathbf{A}$ is the 3-center Coulomb potential (3c-CP) integral tensor,
and $\mat{\Phi}$ and $\mathcal{Q}$ are the collocation matrix and quadrature 
discussed in \cref{sec:xc}.
Inserting \cref{eq:sn_eri} into
\cref{eq:full_k}, we obtain a simple expansion for $\mat{K}$
\cite{friesner1985solution,ringnalda1990pseudospectral,neese2009efficient,laqua20_highly},
\begin{align}
G_{\mu i}      &= \sum_\lambda w_i A_{\mu\lambda i} F_{\lambda i}, \label{eq:gmat}\\
\tilde{K}_{\mu\nu} &= \sum_i G_{\mu i} \Phi_{\nu i}, \label{eq:usym_k}\\
K_{\mu\nu} &\approx \frac{1}{2}\left( \tilde{K}_{\mu\nu} + \tilde{K}_{\nu\mu} \right), \label{eq:sn_k}
\end{align}
where $\mat{F}$ is same intermediate given in \cref{eq:fmat} for the evaluation of $\mathbf{V}^{xc}$. This method for $\mat{K}$ assembly will
be referred to as seminumerical exchange\cite{laqua20_highly} (sn-K) in the following. Apart from the use of common intermediates,
there exists a striking similarity between
\cref{eq:gmat,eq:usym_k,eq:sn_k} and the $\mat{V}^{xc}$ assembly in \cref{eq:vxc_num,eq:zmat,eq:den_grid,eq:fmat}. As has been posited elsewhere
\cite{laqua20_highly,neese2009efficient}, sn-K can reuse many of the same algorithmic primitives as those
used for the evaluation of $\mat{V}^{xc}$. One might even be tempted to
implement \cref{eq:gmat} via BLAS, but we (as have others 
\cite{laqua20_highly,neese2009efficient}) have found this to be inefficient in practice, particularly on GPU architectures. 
We present the
scheme by which we perform the combined evaluation of \cref{eq:3c_grid,eq:gmat} in \cref{sec:3ccp}.

\subsection{Batching and Screening of Numerical Integrals}
\label{sec:batch_screen}

Each of the expressions in \cref{eq:vxc_num,eq:collocation,eq:zmat,eq:exc_grid,eq:den_grid,eq:fmat,eq:gmat,eq:usym_k} is \emph{local} with respect to the numerical
quadrature, 
and thus may be
partitioned along their grid indices ($i$) with final results (e.g., \cref{eq:vxc_num,eq:usym_k}) being recovered as a sum over locally
integrated intermediates. This partitioning has clear ramifications in
the context of parallelism which will be discussed in
\cref{sec:load_balance}. Grid point locality also
allows for the exploitation of the spatial sparsity of 
$\mathcal{B}$ and various operators (e.g. \cref{eq:3c_grid}), which in turn takes the nominal $\mathcal{O}(N_b^2N_g)$ scaling
of these numerical schemes to methods which scale (near-)linearly with
respect to system size.\cite{stratmann96_achieving,neese2009efficient,burow11_linear,laqua20_highly} 

As
the sparsity profiles of spatially adjacent grid points are similar,
it is canonical to \emph{batch} these grid points into subsets 
$\mathcal{Q}_j$ such that $\mathcal{Q} = \bigcup_j\mathcal{Q}_j$ with
$\mathcal{Q}_j\cap\mathcal{Q}_k=\emptyset$ for $j\neq k$. Several approaches for quadrature grid
batching have been suggested in various contexts. \cite{stratmann96_achieving,burow11_linear,laqua20_highly}
In this work, we utilize an octree-based partition scheme
which has been used in previous studies by the authors. \cite{williams20_on} 
In general, the octree approach will yield more spatially
local quadrature batches than other approaches\cite{laqua20_highly}, but will also yield batches
of irregular sizes which potentially leads to load imbalance between tasks. We discuss the implications of this
irregularity in the following.
Given a partitioning $\{\mathcal{Q}_j\}_{j=1}^{N_\mathrm{batch}}$, one may construct sets
of important quantities which are approximately considered 
non-negligible for a particular integrand (e.g., $\mathbf{K}$ or $\mathbf{V}^{xc}$). The simplest of these sets, which is required for both numerical integrands considered in this work, pertains to 
negligible basis functions, 
\begin{equation}\mathcal{B}_j = \{\phi_\mu\,\,\vert\,\,\exists \mathbf{r}_i\in\mathcal{Q}_j\quad\mathrm{s.t.}\quad\vert\phi_\mu(\mathbf{r}_i)\vert\geq \varepsilon_{B} \}, \label{eq:basis_screen}
\end{equation}
where $\varepsilon_B$ is a basis tolerance. 
For GTO bases, \cref{eq:basis_screen} may be quickly determined by encircling each basis function with a sphere of
radius $r^\mathrm{cut}_\mu$ such that  
$\vert\phi_\mu(\mathbf{r}-\mathbf{R}_\mu)\vert< \varepsilon_B$ $\forall\vert\mathbf{r} - \mathbf{R}_\mu\vert > r^\mathrm{cut}_\mu$, 
and only keeping those elements of $\mathcal{B}$ for which
their non-negligible sphere spatially overlaps $\mathcal{Q}_j$\cite{neese2009efficient,stratmann96_achieving,laqua20_highly,williams20_on}.
This process formally scales 
$\mathcal{O}(N_bN_\mathrm{batch})$ as each basis shell must be checked against each $\mathcal{Q}_j$. It is worth noting that in scenarios where the nuclei remain
stationary for several Fock builds, e.g., in a self-consistent field optimization (SCF), $\mathcal{B}_j$ need only be computed once and
reused for subsequent iterations.
This process of basis screening
is the primary mechanism by which sparsity may be exploited in
the numerical integration of $\mathbf{V}^{xc}$\cite{stratmann96_achieving}. For example,
given $\mathcal{B}_j$, 
the compute-intensive \cref{eq:fmat} becomes
\begin{equation}
    F^{xc, (j)}_{\mu i} = \sum_{\nu\in\mathcal{B}_j} D^{xc,(j)}_{\mu\nu}
    \Phi^{(j)}_{\mu i}, \qquad \mu\in\mathcal{B}_j, i\in\mathcal{Q}_j,
\label{eq:block_sparse_fmat_xc}
\end{equation}
where $\mathbf{D}^{xc,(j)}$, $\mathbf{F}^{xc, (j)}$ and $\boldsymbol{\Phi}^{(j)}$ are contiguous batch
local matrices pertaining to $\mathcal{Q}_j$.

The case is significantly different for the sn-K method as basis
function and integral screening alone can only produce $\mathcal{O}(N^2)$ scaling\cite{VRG:kohn:1959:PR}. 
 Several methods
have been suggested for density-dependent screening in the 
sn-K method.\cite{friesner1985solution,ringnalda1990pseudospectral,neese2009efficient,laqua20_highly,helmich2021improved}.
Due to its simple form and demonstrated ability 
to generate sufficient sparsity for near-linear scaling, we adopt the
sn-LinK screening approach of Laqua, \emph{et al}\cite{laqua20_highly} in this
work, though we note that other schemes could be employed with only minor
modifications of the overall algorithm design. In the sn-LinK 
method, a list of basis pairs,
$\mathcal{V}_{j}$, is selected for each 
$\mathcal{Q}_j$ such that 
\begin{align}
    \mathcal{V}_j &= \{(\phi_\mu,\phi_\nu)\quad\vert\quad
    \varepsilon^{E(j)}_{\mu\nu}\geq\varepsilon_E\,\lor\,
    \varepsilon^{K(j)}_{\mu\nu}\geq\varepsilon_K\}, \label{eq:sp_set}
\end{align}
where $\varepsilon_K$ and $\varepsilon_E$ are tolerances which screen
shell-pair contributions to $\mathbf{K}$ and $\mathrm{Tr}[\mathbf{KD}]$ (the exact-exchange energy), respectively, and
\begin{align}
    \varepsilon^{E(j)}_{\mu\nu} &= 
    \tilde F_\mu^{K,(j)} \tilde F_\nu^{K,(j)} \tilde A_{\mu\nu}, \\
    \varepsilon^{K(j)}_{\mu\nu} &= 
    \max\left(\tilde F_\mu^{K,(j)}, \tilde F_\nu^{K,(j)} \right)
    \tilde \phi^{(j)}\tilde A_{\mu\nu}, \\
    \tilde F_\mu^{K,(j)} &= \sum_{\nu\in\mathcal{B}_j} 
    \vert D_{\mu\nu}\vert \tilde\Phi_\nu^{(j)}, \label{eq:exx_fmat_approx}
\end{align}
\begin{equation}
        \tilde A_{\mu\nu} = \max_{i\in\mathcal{Q}_j}\vert A_{\mu\nu i}\vert,\quad
    \Phi_\nu^{(j)} = \max_{i\in\mathcal{Q}_j} \vert\sqrt{w_i}\Phi_{\nu i}\vert,\quad 
    \tilde\phi^{(j)} = \max_{i\in\mathcal{Q}_j} \sqrt{w_i}\sum_{\mu\in\mathcal{B}_j}\vert\Phi_{\mu i}\vert.
\end{equation}
To avoid re-computation of the  3c-CP integrals in the screening procedure, 
$\tilde A_{\mu\nu}$ is approximated with global upper-bound estimates from
the appendix of Ref.\citenum{thompson19_integral} in practice \cite{laqua20_highly}. We refer the reader to Laqua, \emph{et al}.\cite{laqua20_highly}
for further details pertaining to the veracity of this screening
procedure for sn-K. 
Given $\mathcal{V}_j$, block-sparse
expressions similar to \cref{eq:block_sparse_fmat_xc} can be
derived for sn-K, e.g.,
\begin{equation}
    F^{K,(j)}_{\lambda i} = \sum_{\nu\in\mathcal{B}_j} D^{K,(j)}_{\lambda\nu} \Phi_{\nu i}^{(j)},\quad
    G^{K,(j)}_{\mu i} = \sum_{(\cdot,\lambda)\in\mathcal{V}_j} w_iA_{\mu\lambda i}F^{K,(j)}_{\lambda i},\quad \mu,\lambda\in\mathcal{V}_j, 
    i\in\mathcal{Q}_j. \label{eq:batch_local_exx}
\end{equation}
The $(\cdot,\lambda)$-sum pertaining to $G^{K,(j)}_{\mu i}$ indicates
only inclusion of terms for which $\exists \phi_\mu$ such that $(\phi_\mu,\phi_\lambda)\in\mathcal{V}_j$. Both the $\mathcal{B}$-collision detection
and the evaluation of performance critical $\mathcal{V}_j$-intermediates (e.g., \cref{eq:exx_fmat_approx} can be implemented as a single, large dimension-GEMM
over tasks) are executed on the GPU in this work.

There are several key differences in screening processes for sn-K in comparison with the XC integration:
\begin{enumerate}
    \item As this procedure is density-dependent, $\mathcal{V}_j$ must be reevaluated each new density 
    and its cost cannot generally be amortized over e.g., an SCF
    optimization.

    \item While the matrix multiplication in \cref{eq:exx_fmat_approx} can be restricted over $\nu\in\mathcal{B}_j$, the $\mu$-index
    \emph{must} be over the entire basis set $\mathcal{B}$ in order
    to check for all possible non-trivial elements of $\mathcal{V}_j$.
    Therefore, \cref{eq:exx_fmat_approx} formally scales 
    $\mathcal{O}(N_bN_\mathrm{batch})$ assuming $\mathcal{B}_j$ can be 
    screened to an amortized constant.\cite{stratmann96_achieving} While
    this formally scales the same as $\mathcal{B}$-screening, it
    will be accompanied by a much larger prefactor (GEMM vs scalar collision detection).

    \item \cref{eq:sp_set} scales $\mathcal{O}(N_bN_\mathrm{batch})$ as each non-negligible shell-pair (which asymptotically scales $\mathcal{O}(N_b)$) must be checked against every $\mathcal{Q}_{j}$. 

\end{enumerate}
 We discuss the implications of this screening procedure on the overall performance and scalability of numerical integration procedures in \cref{sec:results}.

\subsection{Load Balancing and Parallel Numerical Integration}
\label{sec:load_balance}

In addition to enabling the screening of basis functions and operator integrals,
the locality of the numerical integration procedure considered in this work is also
particularly advantageous for distributed memory implementations \cite{apra20_nwchem,williams20_on,manathunga21_harnessing}. As the batch-local quantities
discussed in the previous subsection are independent, they may be executed concurrently
and assembled \emph{a postiori} via collective reduction operations to form final integrands. 
The general distributed memory scheme explored in this work is a simple three step process (\cref{alg:dist_xc}) that
is the same for both the sn-K and XC integrations\cite{williams20_on}.

\begin{algorithm}[t]
\begin{algorithmic}[1]
\State \textbf{Input:} Density matrix $\mathbf{D}$, basis set $\mathcal{B}$.
\State \textbf{Output:} Desired integrand $\mathbf{X}\in\{\mathbf{V}^{xc},\mathbf{K}\}$
\State $\mathcal{Q}_{local}\leftarrow$ Generate balanced local quadrature batches \Comment{\cref{alg:load_balance}}
\State Preallocate fraction of device memory.
\State Send quadrature independent data (e.g. $\mathbf{D}$) to the device \Comment{cpu/gpu}
\State $\mathbf{X}_{local}\leftarrow\mathbf{0}$ \Comment{gpu}
\While{$\mathcal{Q}_{local}\neq\emptyset$}
   \State $\mathcal{Q}_{batch} \leftarrow$ Subset of $\mathcal{Q}_{local}$
   s.t. $\mathbf{X}$-intermediates saturate device memory. \Comment{cpu}
   \State Send $\mathcal{Q}_{batch}$ data (e.g. $\{\mathcal{B}_j\}_\mathrm{local}$, $\{\mathcal{V}_j\}_\mathrm{local}$) to device \Comment{cpu/gpu}
   \State $\mathcal{Q}_{local} \leftarrow \mathcal{Q}_{local}\setminus\mathcal{Q}_{batch}$ \Comment{cpu}
   \State $\mathbf{X}_{local}\leftarrow\mathbf{X}_{local} + \mathcal{Q}_{local}$ 
   contributions to $\mathbf{X}$ \Comment{gpu}
\EndWhile
\State $\mathbf{X}\leftarrow$\texttt{(All)reduce}($\mathbf{X}_{local}$) \Comment{collective}
\end{algorithmic}
\caption{General scheme for the parallel, GPU-accelerated evaluation of $\mathbf{V}^{xc}$ or $\mathbf{K}$}
\label{alg:dist_xc}
\end{algorithm}

Due to the varying extent to which sparsity can be realized between differing spatial regions,
there is a significant potential for load imbalance in parallel implementations of
molecular integration which rely on grid partitioning.\cite{williams20_on,manathunga20_parallel,manathunga21_harnessing}
The optimal task assignment problem for irregular work is NP-Hard\cite{sanders2019sequential}, but reasonable solutions can be obtained 
using heuristic driven methods. A recent distributed memory
DFT application\cite{manathunga21_harnessing}
has adopted a dynamic load balancing scheme to
address this problem. At large processor counts, dynamic load balancing algorithms
rely on the ability to overlap the costs associated with the generation, assignment,
and communication of tasks with their execution on local processors. Given the
fine-granularity and large number of tasks generated by the octree-batching algorithm
discussed in the previous section, the communication costs associated with the memory
requirement of each task (e.g., $\mathcal{Q}_j$, $\mathcal{B}_j$ and $\mathcal{V}_j$)
far outweigh their associated computational work. 
Instead, we have developed a \emph{static} load balancing procedure for 
molecular integration which allows for the \emph{a priori} distribution of
work without the need to carefully balance the overlap of local work and 
task communication. In this static load balancing scheme, a 
heuristic cost, $W_j$,
is associated with each task and is subsequently assigned to a rank
via a greedy algorithm illustrated in \cref{alg:load_balance}. For the XC
integration, the work heuristic is given by\cite{williams20_on}
\begin{equation}
    W_j = \left(\vert\mathcal{B}_j\vert\times(9 + 2\vert\mathcal{B}_j\vert) + N_A^2 + 3 \right)\times \vert\mathcal{Q}_j\vert.\label{eq:work_heuristic}
\end{equation}
\Cref{alg:load_balance} is a replicated procedure which allows for near optimal task
assignment given work heuristics which properly rank the relative computational 
costs between different tasks. As such, while being communication-free, it constitutes a scaling bottleneck due
to Amdahl's law. Similar algorithms based on prefix-sums
have been suggested for the distributed generation of tasks\cite{sanders2019sequential} However, we have found
that because \cref{alg:load_balance} is able to globally optimize task assignment (via the sort of all tasks on $W_j$), 
it generally is able to obtain a more optimal task partitioning than those based
on prefix sums. Due to the fact that $\mathcal{B}_j$ in particular can be computed
once-per-nuclear configuration, the scaling bottleneck associated with the replicated
nature of \cref{alg:load_balance} can be amortized over several Fock builds for the
XC integration.

\begin{algorithm}[t]
\begin{algorithmic}[1]
    \State \textbf{Input:} Quadrature batches $\{\mathcal{Q}_j\}_{j=1}^{N_\mathrm{batch}}$
    \State \textbf{Output:} Local quadrature batches $\mathcal{Q}_\mathrm{local}\subset\{\mathcal{Q}_j\} $
    \State \texttt{world\_size} $\leftarrow$ Number of ranks
    \State \texttt{world\_rank} $\leftarrow$ Index of this rank
    \State $\mathcal{T} \leftarrow [\,]$
    \For{$j\in [1,N_\mathrm{batch}]$}
      \State $\mathcal{T}_j = (\mathcal{Q}_j, W_j)$\Comment{\cref{eq:work_heuristic}}
    \EndFor
    \State $\mathcal{T} \leftarrow$ Sort $\mathcal{T}$ by $W_j$
    \State $\mathcal{G}\leftarrow$ \texttt{zeros}( \texttt{world\_size} )
    \State $\mathcal{Q}_\mathrm{local}\leftarrow[\,]$
    \For{$j\in [1,N_\mathrm{batch}]$}
        \State $I\leftarrow\arg\min_J\mathcal{G}_J$ \Comment{global}
        \State $\mathcal{G}_I\leftarrow\mathcal{G}_I + W_j$ \Comment{global}
        \If{$I=\texttt{world\_rank}$}
          \State $\mathcal{Q}_\mathrm{local} \leftarrow \mathcal{Q}_\mathrm{local} \cup \mathcal{Q}_j$ \Comment{local}
        \EndIf
    \EndFor
    \State \Return $\mathcal{Q}_\mathrm{local}$
\end{algorithmic}
\caption{Replicated load balancing algorithm for distributed memory molecular integration.}
\label{alg:load_balance}
\end{algorithm}

For the sn-K integration, the problem becomes slightly more difficult. While it would
be possible to generalize \cref{eq:work_heuristic} to account for 
$\vert\mathcal{V}_j\vert$, replication of \cref{eq:sp_set,eq:exx_fmat_approx} on each
processor for all tasks would be a considerable bottleneck due to the steep scaling
and prefactors associated with sn-K screening discussed in the previous subsection. 
Given an initial work partitioning generated by e.g., \cref{eq:work_heuristic}, 
$\mathcal{V}_j$ may be computed locally, and could in principle be re-balanced according
to the distributed prefix-sum algorithms previously mentioned. Apart from generally
obtaining less-optimal work partitions than \cref{alg:load_balance}, for hybrid
DFT simulations which require the evaluation of both $\mathbf{V}^{xc}$ and $\mathbf{K}$,
this procedure would require rebalancing the work distribution at every SCF step, which
would in turn incur significant communication overhead. Instead, we have chosen to
reuse the same task distribution for both sn-K and the XC integration based on
\cref{alg:load_balance,eq:work_heuristic}.  We examine the veracity of this reuse 
to yield balanced work for both sn-K and the XC integration in \cref{sec:results}.

If screened sufficiently well, the dimensions of the packed batch-local
sub-matrices in \cref{eq:collocation,eq:zmat} will be small which means
that the GEMM operations in e.g., \cref{eq:block_sparse_fmat_xc,eq:batch_local_exx} will be of 
low dimension.
While this screening leads to a significant decrease in the 
required computational work, any particular GEMM operation (or
other kernel invocation)
will not constitute enough work to occupy the resources of a GPU.
While it is possible to concurrently execute GPU kernels via 
streaming (as has been explored in other works for GPU-accelerated sn-K integration
\cite{laqua20_highly}), the large number of kernels invoked for large systems partitioned by the octree-method
would incur significant kernel launch overhead. As has been demonstrated
in previous studies regarding the XC integration\cite{williams20_on,williams21_achieving}, the use of 
\emph{batched} kernels to concurrently execute logically identical tasks
can lead to large performance improvements over concurrent stream
injection for fine task granularity. The challenge for DFT applications is in that the dimensions of
$\mathcal{B}_j$ and $\mathcal{V}_j$ can vary drastically in different spatial
regions. For the evaluation of batched GEMM operations (e.g., \cref{eq:block_sparse_fmat_xc,eq:batch_local_exx} batched over $j$ indices),
standard solutions to the problem exist in community GPU linear algebra
libraries in the form of variable-sized batched BLAS.\cite{haidar15_batched,abdelfattah16_performance}

For sn-K, the batched evaluation of 
\cref{eq:usym_k} is a straight-forward application of variable-sized batched GEMM as was
applied to the formation of \cref{eq:block_sparse_fmat_xc} for the XC integration.
Batching of \cref{eq:gmat} can be achieved hierarchically over three dimensions
(1) $\{\mathcal{Q}_j\}$, (2) $(\mu,\nu)\in\mathcal{V}_j$ within $\mathcal{Q}_j$,
and at the lowest level (3) $i\in\mathcal{Q}_j$. The kernel-level details for (3) are
given in \cref{sec:3ccp}. For any batched kernel algorithm, the natural challenge
is to decide which task to batch together to execute concurrently on the device.
For the algorithms presented here, we make the simple choice of executing as many
tasks as will fit into device memory (\cref{alg:dist_xc}). While this choice does 
mitigate kernel launch overhead and allows for the overlap of host data manipulations 
with GPU activity, it also introduces significant pressure on the GPU scheduler
to effectively execute large numbers of independent tasks with variable amounts of 
work. However, we have seen that for DFT applications that this device-saturation approach
is capable of yielding excellent utilization of GPU resources on an array of modern
GPU hardware.\cite{williams20_on,williams21_achieving}.

In the large processor limit, the reduction step
poses a considerable bottleneck for large $N_b$ as the message size for the \texttt{(All)reduce} operation scales $\mathcal{O}(N_b^2)$.
While standard MPI collectives are often sufficient for small $N_b$, we have found that for NVIDIA hardware,
the NVIDIA Collective Communication Library (\NCCL)\cite{nccl} all-reduce primitive can yield significant strong scaling improvements for large $N_b$. \NCCL provides collective primitives with an API similar to that of MPI with the addition of a CUDA stream argument which allows for its injection into asynchronous workflows. The all-reduce algorithm and exact choice of communication routes are selected by \NCCL at runtime based on the system topology allowing for optimal performance on a wide-range of hardware configurations and message sizes.  We will examine the comparison of MPI and \NCCL collectives for this application in \cref{sec:results}.

\section{Numerical Results}
\label{sec:results}

The methods presented in this work are made publicly available as a part of the open-source
\texttt{GauXC} library for exascale
Gaussian basis DFT (XC and sn-K)\cite{petrone2018efficient,williams20_on,williams21_achieving} and the \texttt{LibintX} library for  Gaussian AO integrals (DF-J-engine).\cite{doi:10.1021/acs.jctc.2c00995} Each
of these libraries were developed to be modular, reusable GPU-based components under the NWChemEx Exascale Computing Project and are freely available on GitHub\cite{gauxc,libintx} for use in other electronic structure packages. 
As present, these methods have been integrated into the \texttt{NWChemEx}\cite{kowalski21_from} and \texttt{MPQC}\cite{VRG:peng:2020:JCP} 
computational chemistry software packages.
In the present study, all numerical results were obtained using \texttt{MPQC}; however, as the focus of this work is on the parallel construction of hybrid DFT Fock matrices, the performance characterizations presented here are expected to be representative of any software integration involving these libraries.
All numerical
experiments were carried out on the GPU partition of the 
Perlmutter (PM) supercomputer at the National Energy Research Scientific Computing Center (NERSC). Each PM GPU node has 4 NVIDIA A100 GPUs (40GB HBM2e RAM) and 1 AMD EPYC 7763 CPU (64 cores @ 3.5GHz). The GPUs within a PM node are connected via NVLink and are connected to the CPU via PCI-e. Internode communication is facilitated through the HPE Slingshot 11 interconnect.
All numerical experiments were configured with 1 MPI
rank per GPU (4 ranks per node, 16 threads per rank) and the threads
of each MPI rank were bound to respective NUMA domains (of which there are 4 per PM node). GPU Batched BLAS operations for the sn-K and XC integrations
were provided by the MAGMA library \cite{haidar15_batched,abdelfattah16_performance}. 

\begin{table}[]
    \centering
    \begin{tabular}{l|ccc}
         Molecule &  $N_A$ & $N_b$  & $N_\mathrm{aux}$  \\
         \hline\hline
         Taxol     & 113    & 1,032 & 3,599\\
         Olestra   & 453    & 3,181 & 11,633\\
         Crambin   & 642    & 5,559 & 19,500\\
         Ubiquitin & 1,231 & 10,292 & 36,419
    \end{tabular}
    \caption{Representative systems considered in this study. $N_A$ is the number of atoms, $N_b$ is given for Cartesian 6-31G(d) and
    $N_\mathrm{aux}$ for def2-tzvp-j.}
    \label{tbl:mol}
\end{table}

\begin{table}[]
    \centering
    \begin{tabular}{l|cc}
         Grid Name & $N_\mathrm{rad}$ & $N_\mathrm{ang}$ \\
         \hline\hline
         \texttt{Grid1} & 50 & 302 \\
         \texttt{Grid2} & 30 & 110
    \end{tabular}
    \caption{Spherical atomic quadratures consisting of $N_\mathrm{rad}$ radial point and $N_\mathrm{ang}$ angular points used in this work. Angular resolution has been pruned according to the scheme of Ref.\citenum{treutler95_efficient}}
    \label{tab:grids}
\end{table}

As have been
utilized in other studies \cite{williams20_on,manathunga21_harnessing}, we have included experiments with
4 test molecules, the specifics of which can be found in \cref{tbl:mol}. The geometries
for Taxol, Olestra and Ubiquitin were taken from Ref.\citenum{williams20_on} and the Crambin geometry was taken from Ref.\citenum{manathunga21_harnessing}. Each of these geometries can be found in the Supplemental Information. 
All experiments utilize the Pople 6-31G(d) basis set\cite{ditchfield1971a,francl1982a,gordon1982a} with Cartesian $d$-functions for $\mathcal{B}$, def2-tzvp-j fitting basis for DF-J\cite{weigend1998a}, and the
PBE0 hybrid XC functional \cite{pbe0}. GPU evaluation of the XC functional was performed
using the open source \texttt{ExchCXX} library\cite{exchcxx}. All calculations were performed with
$\varepsilon_B=\varepsilon_E=\varepsilon_K=10^{-9}$  and with a maximum grid partition size of 4096.
Evaluation of 3-center integrals in \cref{eq:V_df,eq:J_df} is screened using user-provided screener that is provided by the user via the \texttt{LibintX} API. All computational experiments used the default (Schwarz) screener of 3-center AO integrals which omitted evaluation of integral sets below the 64-bit floating point epsilon ($\approx 2*10^{-16}$).

\begin{figure}[!t]
\centering
\begin{minipage}{0.49\textwidth}
\importlocalfigure{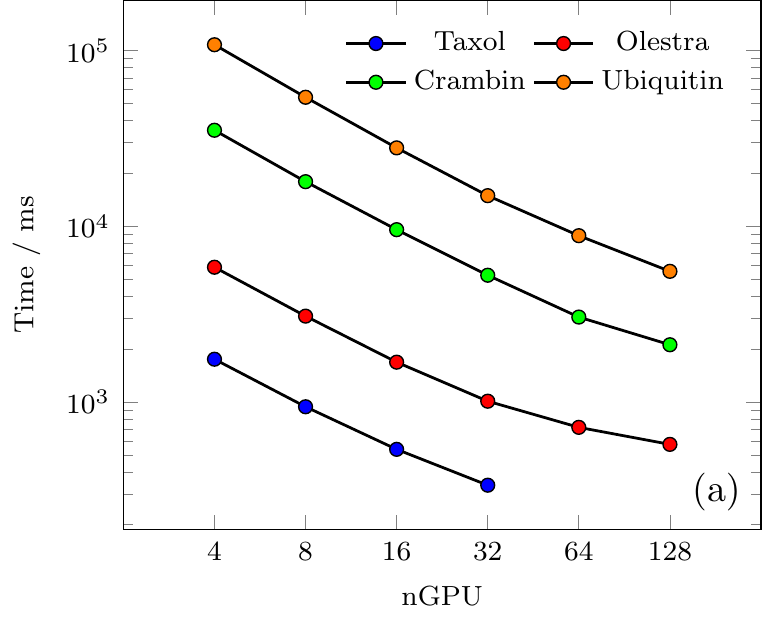}
\end{minipage}~
\begin{minipage}{0.49\textwidth}
\importlocalfigure{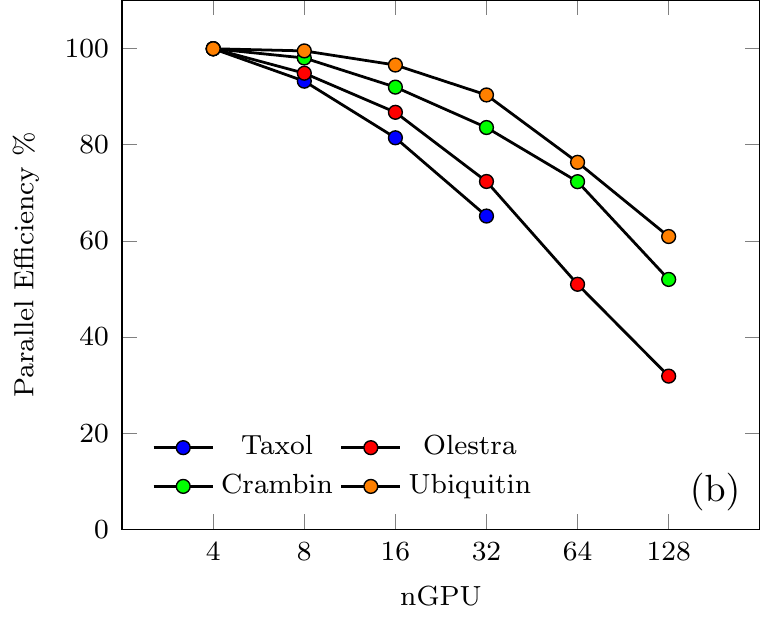}
\end{minipage}~
\caption{Strong scaling of distributed memory Fock algorithm for \texttt{Grid1}. (a) Presents the overall timings and (b) illustrates the parallel efficiency relative to single-node performance}
\label{fig:fock_scaling_gm5}
\end{figure}

\begin{figure}[!t]
\centering
\begin{minipage}{0.49\textwidth}
\importlocalfigure{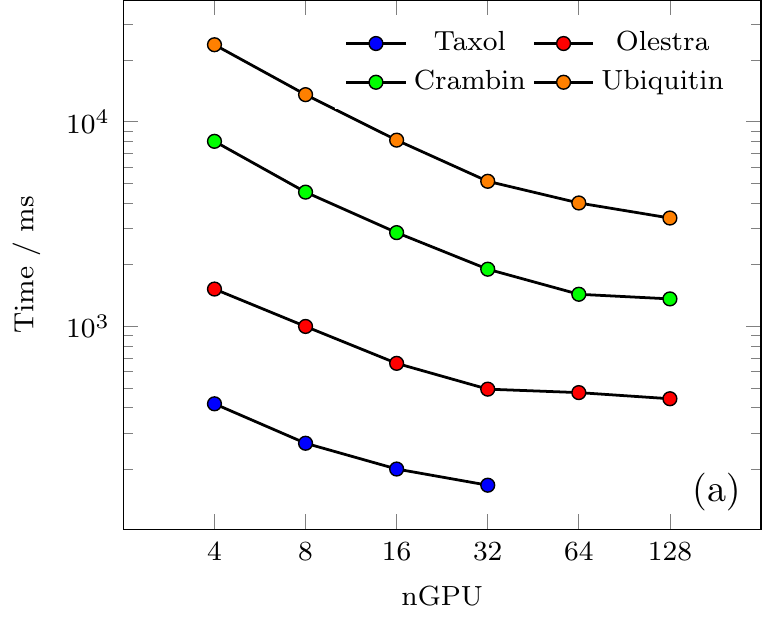}
\end{minipage}~
\begin{minipage}{0.49\textwidth}
\importlocalfigure{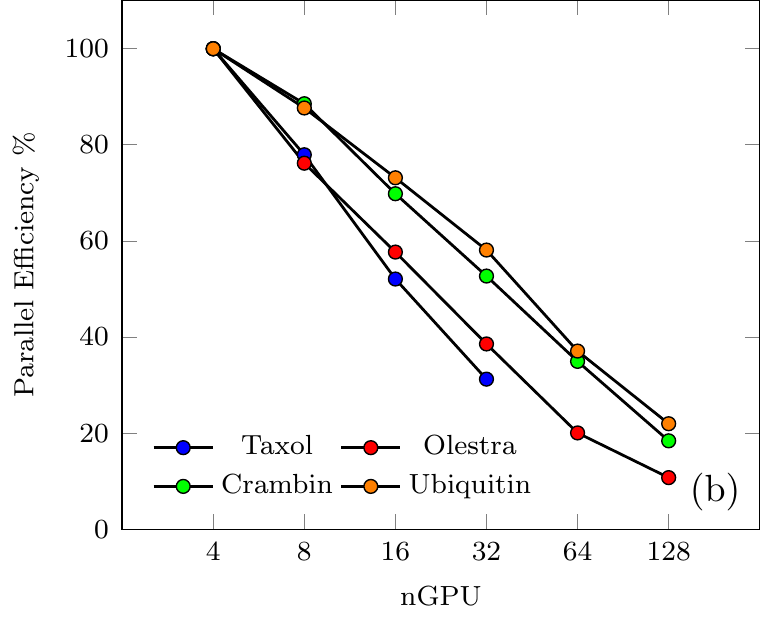}
\end{minipage}
\caption{Strong scaling of distributed memory Fock algorithm for \texttt{Grid2}. (a) Presents the overall timings and (b) illustrates the parallel efficiency relative to single-node performance}
\label{fig:fock_scaling_gm3}
\end{figure}

\Cref{fig:fock_scaling_gm5,fig:fock_scaling_gm3} show the strong scaling 
performance and parallel efficiency (PE) of the overall Fock matrix construction for two different integration
grids (summarized in \cref{tab:grids}). The same grid is used for both XC and sn-K. PE is measured relative to single-node (4 GPU) performance
and all results are presented with the \texttt{NCCL} reduction optimization discussed in \cref{sec:load_balance}. 
For the larger grid (\texttt{Grid1}), we see that excellent PE ($>80\%$) is achieved for all considered problems out to 16 GPUs and 
maintains a very reasonable PE ($>70\%$) out to 32 GPUs except for Taxol which exhibits 64\%. 
This result is comparable with the achieved PE in other
recent distribution memory Fock algorithms at similar GPU counts.\cite{manathunga21_harnessing,barca21_faster}
For Crambin, 
$>70\%$ PE is maintained out to 64 GPUs and achieves 52\% PE at 128 GPUs. For the largest problem (Ubiquitin), $>$90\% PE is maintained out to 32 GPUs, $>75\%$ PE out to 64 GPUs and achieves 61\% PE at 
128 GPUs. For small problems (Taxol and Olestra) and the smaller grid (\texttt{Grid2}), 
strong scaling stagnation is encountered immediately.
Apart from the self-evident 
increase in available local work exhibited by the larger test problems and grids, 
the smaller test problems expose bottlenecks (such as communication and load imbalance) in the relative performance and scalability 
of the individual Fock matrix components. 
The minimum time to solution for these problems, including individual component timings, are accounted in \cref{tbl:tts}.

\begin{table}[]
    \centering
    \begin{tabular}{l|c|ccc|ccc}
         \multicolumn{1}{c}{}       &  \multicolumn{1}{c}{}      & \multicolumn{3}{c}{\texttt{Grid1}} & \multicolumn{3}{c}{\texttt{Grid2}} \\
         Molecule &  DF-J & XC & sn-K & Fock & XC & sn-K & Fock \\
         \hline \hline
         Taxol     & 63.2    & 69.4     & 107.5    & 240.0     & 95.1     & 256.9     & 423.6   \\
         Olestra   & 113.5   & 320.5    & 438.0    & 872.0     & 363.6    & 565.9     & 1,043.3 \\
         Crambin   & 177.8   & 968.6    & 1,512.6  & 2,659.0   & 1,043.4  & 2,226.4   & 3,448.8 \\
         Ubiquitin & 378.8   & 2,268.0  & 3,674.0  & 6,320.8   & 2,443.1  & 5,653.1   & 8,465.6
    \end{tabular}
    \caption{Minimum time to solution for considered systems on 128 NVIDIA A100 GPUs. All times are given in millisections (ms)}
    \label{tbl:tts}
\end{table}

\begin{figure}[tbh]
\centering
\begin{minipage}{0.49\textwidth}
\importlocalfigure{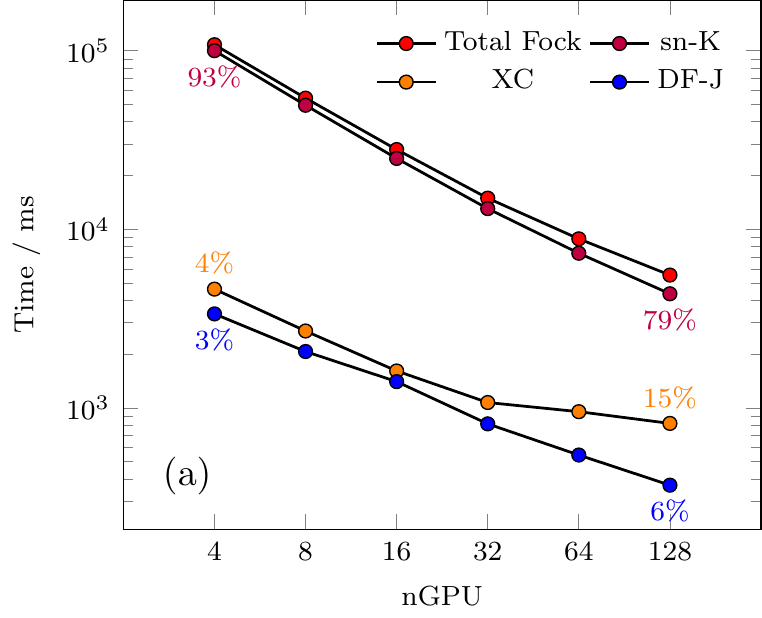}
\end{minipage}~
\begin{minipage}{0.49\textwidth}
\importlocalfigure{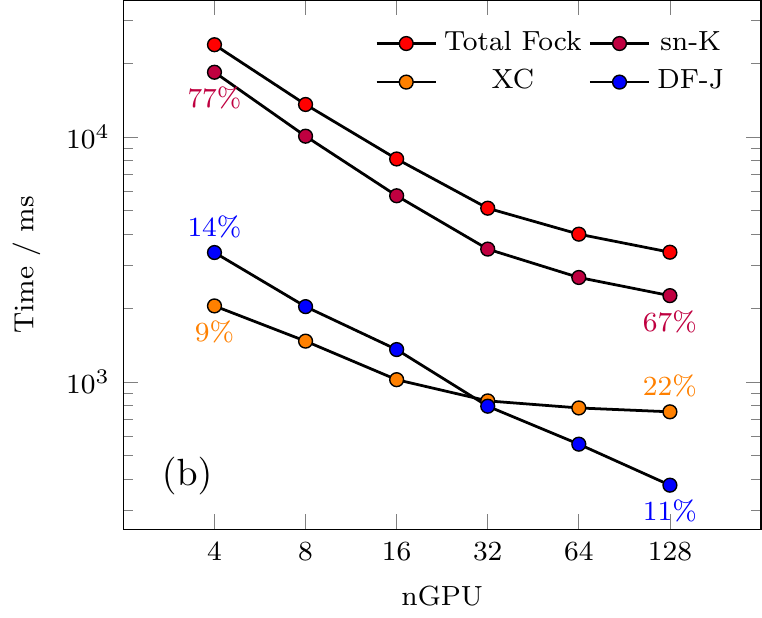}
\end{minipage}
\caption{Strong scaling of individual Fock matrix components for Ubiquitin 6-31G(d)/def2-tzvp-j/PBE0 using (a) \texttt{Grid1} and (b) \texttt{Grid2} for the XC and sn-K integration. Annotations depict the relative contribution of each component to the overall Fock timings.}
\label{fig:ubi_fock_scaling}
\end{figure}

\begin{figure}[tbh]
\centering
\begin{minipage}{0.49\textwidth}
\importlocalfigure{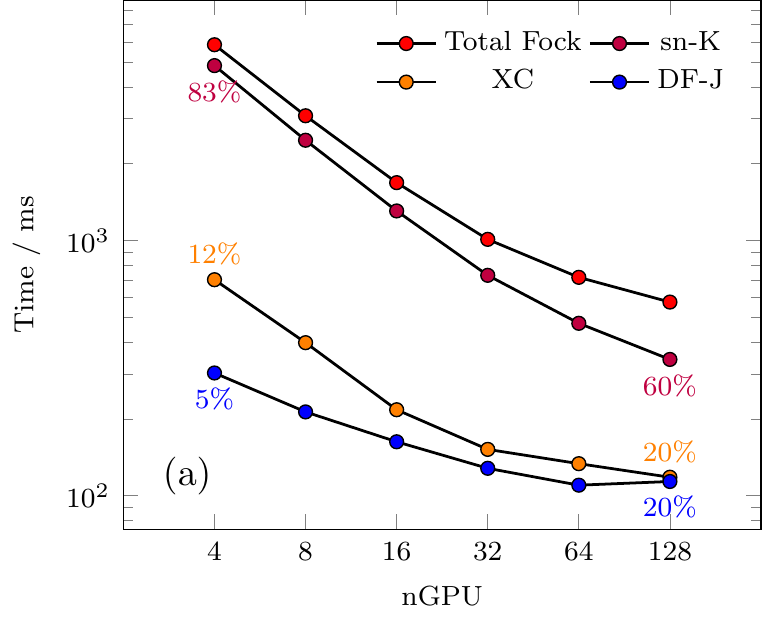}
\end{minipage}~
\begin{minipage}{0.49\textwidth}
\importlocalfigure{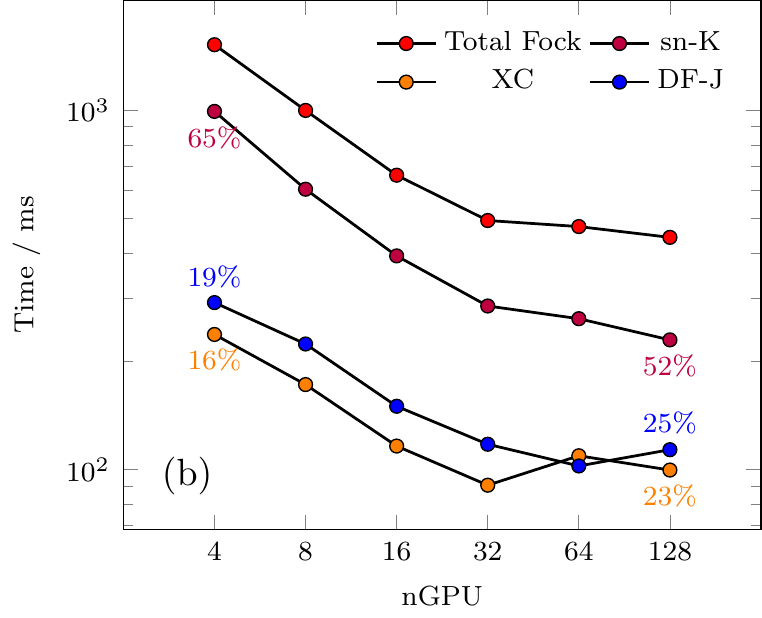}
\end{minipage}
\caption{Strong scaling of individual Fock matrix components for Olestra 6-31G(d)/def2-tzvp-j/PBE0 using (a) \texttt{Grid1} and (b) \texttt{Grid2} for the XC and sn-K integration. Annotations depict the relative contribution of each component to the overall Fock timings.}
\label{fig:ole_fock_scaling}
\end{figure}

\Cref{fig:ubi_fock_scaling,fig:ole_fock_scaling} illustrate the strong scaling of the $\mathbf{J}$, $\mathbf{K}$ (sn-K)
and $\mathbf{V}^{xc}$ components relative to the total Fock formation for the Olestra
and Ubiquitin test cases at various grid sizes. The textual annotations denote the
wall-time percentage of each component relative to the overall Fock build. For the Ubiquitin 
test case (\cref{fig:ubi_fock_scaling}), sn-K dominates the overall Fock build by a large margin at all GPU counts, and the strong scaling 
of the Fock build is virtually identical to that of sn-K (as illustrated by the parallelity
of the scaling plots). As such, the overall scaling behaviour of the Fock build is insensitive
to the scaling behaviours of $\mathbf{J}$ and XC. 
The opposite case is illustrated for Olestra (\cref{fig:ole_fock_scaling}), where the $\mathbf{J}$-build and XC integration
constitute a much more considerable portion of the overall Fock build than was exhibited for
Ubiquitin, particularly at large GPU counts. As such, we can conclude that the scaling behaviours of $\mathbf{J}$ and XC have a much more measurable
impact on the overall strong scaling of the Fock build for smaller systems.

\begin{figure}[b]
\begin{minipage}{0.33\textwidth}
\importlocalfigure{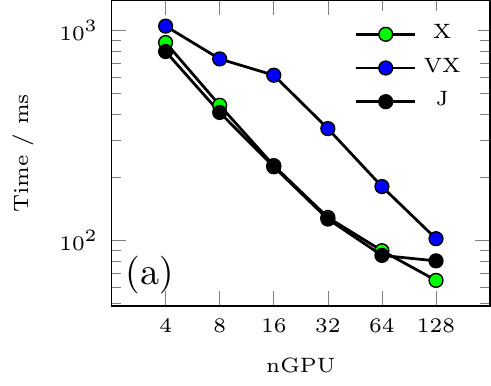}
\end{minipage}~
\begin{minipage}{0.33\textwidth}
\importlocalfigure{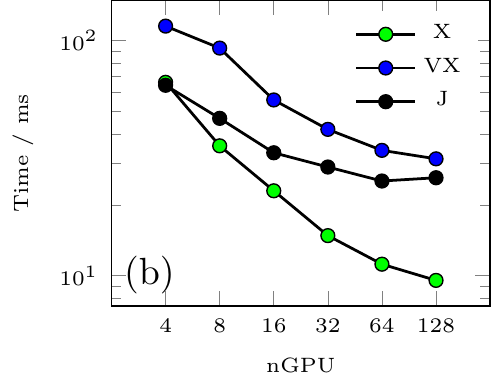}
\end{minipage}~
\begin{minipage}{0.33\textwidth}
\importlocalfigure{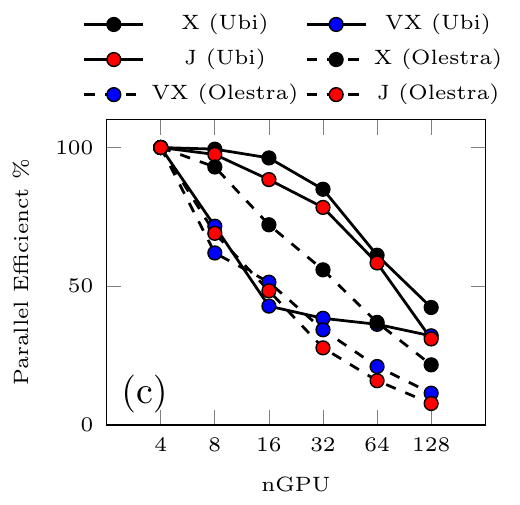}
\end{minipage}
\caption{Component strong scaling of the distributed DF-J algorithm presented in \cref{sec:rij} for (a) Ubiquitin and (b) Olestra 6-31G(d)/def2-tzvp-j. (c) Illustrates the parallel efficiencies
of these components for both problems.}
\label{fig:j_scaling}
\end{figure}

\begin{figure}[t]
\centering
\begin{minipage}{0.33\textwidth}
\importlocalfigure{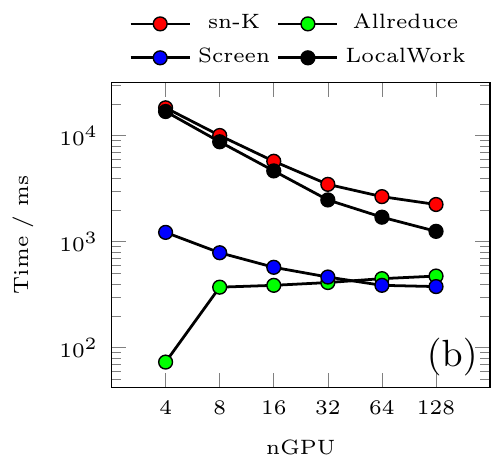}
\end{minipage}~
\begin{minipage}{0.33\textwidth}
\importlocalfigure{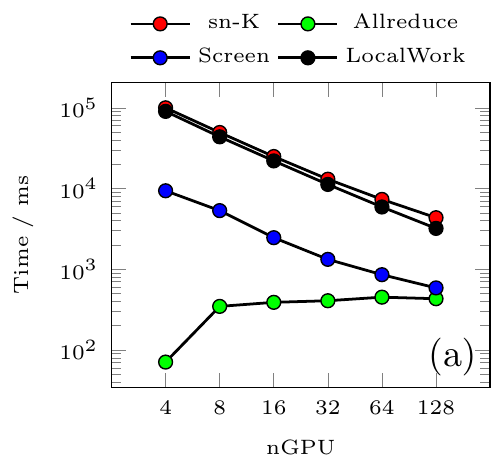}
\end{minipage}~
\begin{minipage}{0.33\textwidth}
\importlocalfigure{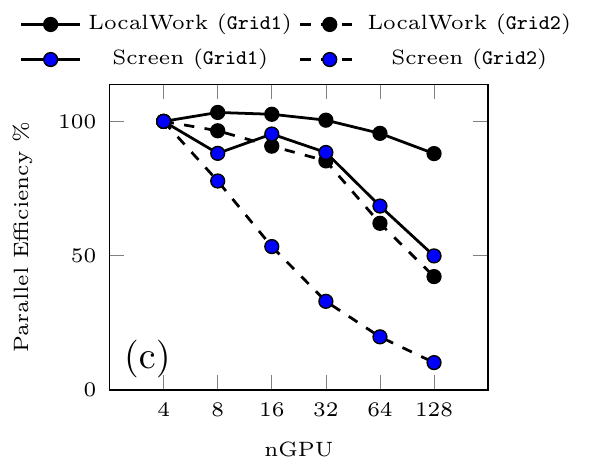}
\end{minipage}
\caption{Component strong scaling of the distributed sn-K algorithm
presented in \cref{alg:load_balance} for Ubiquitin 6-31G(d) using (a) \texttt{Grid1} and (b) \texttt{Grid2} numerical
quadratures. (c) Illustrates the parallel efficiency of these components for both problems.}
\label{fig:snk_scaling}
\end{figure}

The performance and scaling behaviour of the XC integration via the algorithms
presented in this work (modulo the \texttt{NCCL} reduction) have been explored
in other works.\cite{williams20_on,williams21_achieving}~In \cref{fig:j_scaling}
we examine the scalability of the 3 primary components of the DF-J algorithm:
\texttt{DF-V} (\cref{eq:V_df}), \texttt{DF-D} (\cref{eq:P_df}) and  \texttt{DF-J} (\cref{eq:J_df}).
For Ubiquitin, the integral-driven \texttt{DF-V} and \texttt{DF-J}  exhibit $>65\%$
PE for up to 32 GPUs where as the scaling of the TiledArray GEMV in \texttt{DF-D} stagnates immediately.
Lack of scaling in the latter can be attributed to the lack of computational intensity 
exhibited by distributed GEMV.
For Olestra, \texttt{DF-D} and \texttt{DF-J} exhibit immediate scaling stagnation while \texttt{DF-V} 
demonstrates $>50\%$ PE for up to 32 GPUs.

For the sn-K
algorithm (\cref{fig:snk_scaling}), we examine the strong scaling of the local integration (\texttt{LocalWork}),
shell-pair screening (\texttt{Screen}), and collective reduction
(\texttt{Allreduce}) for both grid sizes. Host-to-device and device-to-host
data movement timings are included in the \texttt{LocalWork} timings. For \texttt{Grid1} the overall sn-K integration is dominated by the local integration at
all considered GPU counts. However, for \texttt{Grid2}, the slight deviation
of the sn-K scaling from the \texttt{LocalWork} scaling can be attributed to
the growing importance of the reduction and screening operations in the strong
scaling limit. 
In \cref{fig:snk_scaling}(c), we see that for \texttt{Grid1}, the \texttt{LocalWork} timings exhibit
near perfect strong scaling $>95\%$ out to 64 GPUs and degrades only to 88\% at 128 GPUs.
The \texttt{LocalWork} timings for \texttt{Grid2} and the \texttt{Screen} timings \texttt{Grid1} exhibit
reasonable scalability ($>60\%$) out to 32 GPUs but quickly stagnate thereafter. 
These results indicate that the reuse of the XC scheduling heuristic discussed in \cref{sec:load_balance} 
is viable for the \emph{a priori} scheduling of sn-K integration tasks given enough distributable work.

While remaining communication free, it is important to note that the chosen work
distribution scheme for sn-K  requires the reevaluation of \cref{eq:sp_set,eq:exx_fmat_approx}
for $\mathcal{Q}_\mathrm{local}$ at each SCF step. Due to the fact that the scaling
of each of these terms is linear in $N_g$, it is expected that the work associated
with sn-K screening will decrease as the quadrature batches are distributed amongst
independent ranks. However, in the large processor limit where there are 
very few tasks per rank, there remains a large prefactor associated with the 
formation of $\mathcal{V}_j$ which scales $\mathcal{O}(N_b)$. As such
the scaling stagnation of the \texttt{Screen} timings can be attributed to the $\mathcal{O}(N_b)$ prefactor
of the screening procedure overtaking the scalable distribution
the quadrature batches in the 
large processor limit.

\begin{figure}
\centering
\begin{minipage}{0.45\textwidth}
\importlocalfigure{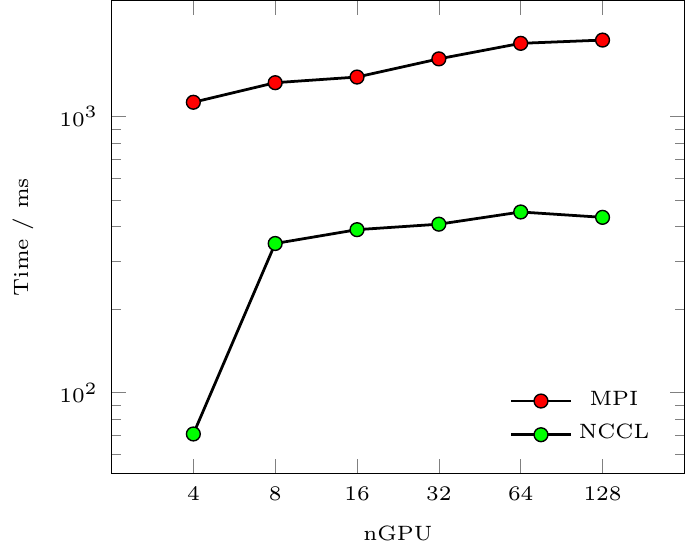}
\end{minipage}~
\begin{minipage}{0.45\textwidth}
\importlocalfigure{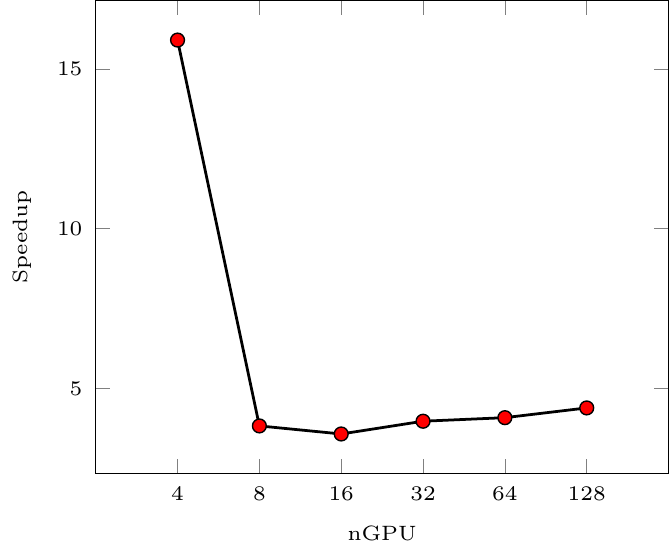}
\end{minipage}~
\caption{Comparison of Cray MPI and \texttt{NCCL} \texttt{Allreduce} performance for Ubiqutin 6-31G(d). 
(a) Shows the strong scaling of \texttt{Allreduce} for these two libraries and (b) illustrates the speedup
of \texttt{NCCL} over Cray MPI.}
\label{fig:nccl_v_mpi}
\end{figure}

It is expected that the cost of the reduction operation will grow with growing
processor counts as the connectivity of both \texttt{Allreduce} communication
graphs grows logarithmically with the number of processors. To demonstrate 
the performance of the \texttt{NCCL} reduction optimization, we compare the Ubiquitin XC/sn-K reduction  with \texttt{MPI\_Allreduce} from the Cray MPI library
in \cref{fig:nccl_v_mpi}. As we can see, \texttt{NCCL} improves the overall communication
cost by about a factor of 4 at all GPU counts, except for 1 node at which the speedup is 15. 
The reason for this speed up can be attributed to the fact that the 4 GPUs on a PM are
connected via NVLink, which allows for a very  fast intranode reduction to take place 
prior to triggering internode communication over the Slingshot interconnect. This fast intranode reduction is clearly
present in the 1 node case where no internode communication is triggered.
When using the MPI reduction primitives, sn-K and XC exhibit a 20\% reduction
in parallel efficiency due to the growing relative communication cost in the strong scaling limit.

The consistent stagnation of strong scaling behaviour as 
problem size (e.g. $N_g$ and/or $N_b$) decreases can be 
attributed to two
primary factors (1) communication overhead and (2) a lack of
divisible work to be distributed amongst independent ranks leading to load imbalance in the large processor limit. The communication overhead (\cref{fig:nccl_v_mpi}) represents a slow growing, nearly constant
cost with processor count, which means that even in the case of perfect strong scaling for the local work
of the Fock build, communication will eventually
dominate the strong scaling behaviour. While these scaling inefficiencies 
are considerable in the large processor limit, It is important to
contextualize the scaling behaviour of these smaller problems 
in reference to their overall wall time at this scale (\cref{tbl:tts}). For example,
while the DF-J algorithm becomes a considerable wall time contribution for the smaller problems in the strong scaling
limit, the overall wall time for this operation is $<$0.4s
for all problems considered. Further, the 128 GPU timing for 
sn-K is under $<=$0.6s for Olesta and Taxol, $<$3s for Crambin and $<$ 6s 
for Ubiquitin. Due to the small timing margins at this scale, it is unlikely that further
algorithmic optimization would overcome inherent
bottlenecks to significantly change the presented scaling results.

\section{Conclusions}
\label{sec:conclusions}

In this work, we have presented a set of GPU accelerated, distributed memory algorithms for the evaluation of the performance-critical Coulomb and exact-exchange matrices for Gaussian basis DFT.
For the Coulomb matrix, we developed a DF-based J-engine implementation capable of efficient execution on distributed memory heterogeneous platforms. This work extends our recent developments of algorithms for evaluation of 3-center integrals on accelerated architectures.\cite{doi:10.1021/acs.jctc.2c00995} For the exact-exchange matrix, we have developed a distributed memory sn-K algorithm which extends recently developed numerical integration 
methods for the Gaussian basis XC potential\cite{williams20_on}. These methods were
implemented in the open-source \texttt{LibintX} and \texttt{GauXC} libraries and 
are publicly available on GitHub\cite{libintx,gauxc}. We have demonstrated the absolute and strong scaling performance of the presented algorithms for a representative
set of molecular systems out to 128 NVIDIA A100 GPUs on the Perlmutter supercomputer. For the
largest problem considered (Ubiquitin), total Fock construction via our methods exhibited $>90\%$ parallel
efficiency out to 32 GPUs and achieved 61\% efficiency out to 128 GPUs for a total
Fock matrix duration of approximately 8.5s.
While degraded
strong scalability was exhibited for the smaller problems considered due to
communication overhead and load imbalance, the magnitude of their effects 
on overall scalability are magnified relative to low execution time 
of these operations and are small on an absolute scale.

While results of the present work indicate a promising future for distributed
memory, GPU accelerated algorithms for Gaussian basis DFT, there remain several
areas for exploration in future work.
The efficient evaluation of the Coulomb potential described here can be combined straightforwardly with fast $[\mathcal{O}(N)]$ approaches for evaluation of Coulomb potential, such as the fast multipole method (FMM) that can be applied in both non-periodic\cite{VRG:white:1994:CPL} and periodic\cite{VRG:challacombe:1997:JCP,VRG:kudin:1998:CPL} settings, as well as the Ewald summation\cite{VRG:ewald:1921:AP} for periodic systems. 
While the seminumerical method was only applied to the problem of exact exchange
in this work, similar approaches have been developed for treatment of
correlated many-body methods as well\cite{dutta2018accelerating,murphy1995pseudospectral}. Pursuance of many-body extensions
to the presently developed distributed memory sn-K method will be
the subject of future work by the authors.
Finally, as has recently been demonstrated for the XC integration\cite{williams21_achieving},
the modular nature \texttt{GauXC} library allows for rapid development of
performance portable DFT methods by separating the implementation details
of performance critical kernels from their inclusion in high-level workflows,
thereby exposing the highest potential and flexibility for targeting current
an emerging accelerator hardware with minimal developer effort.
The pursuance of performance portable DF-J and sn-K algorithms will also
be pursued by the authors in upcoming work.

\begin{acknowledgments}
The authors thank Norm M. Tubman for proofreading the draft of this manuscipt and providing useful feedback. This research was supported by the Exascale Computing Project (17-SC-20-SC), a
collaborative effort of the U.S. Department of Energy Office of Science and the
National Nuclear Security Administration.
This research used resources of the National Energy Research Scientific
Computing Center (NERSC), a U.S. Department of Energy Office of Science User
Facility located at Lawrence Berkeley National Laboratory, operated under
Contract No. DE-AC02-05CH11231 using NERSC award ASCR-ERCAP-M3946.
\end{acknowledgments}

\newpage

\appendix
\section{Three-Center Coulomb Potential Integrals}
\label{sec:3ccp}

In this section, we present our approach for the direct evaluation of the contributions arising from
the 3c-CP integrals to the sn-K intermediate 
$\mathbf{G}$ (\cref{eq:gmat}). As opposed to the McMurchie-Davidson approach 
taken for the DF-$J$-engine in \cref{sec:rij}, we have adopted the Obara-Saika GTO integral recursions\cite{obara1986efficient, gill1989efficient} for the 
evaluation of $\mathbf{G}$. We refer the reader for 
\cref{sec:rij} for notation regarding basis functions and integrals used in
this section.

For a particular shell pair $\phi_\mathbf{a},\phi_\mathbf{b}\in\mathcal{B}$, 
and grid point $\mathbf{r}_i\in\mathcal{Q}$, we define
the auxiliary integrals, $\Theta^n_\mathbf{a}$ and $\Xi_{\mathbf{ab}}$ where
\begin{align}
    \Theta^n_{\mathbf{a}+1_q} &= (\mathbf{R}_p - \mathbf{R}_\mathbf{a})_q \Theta^n_{\mathbf{a}} - (\mathbf{P} - \mathbf{r}_i)_q\Theta^{n+1}_\mathbf{a} +
    \frac{a_q}{2(\zeta_\mathbf{a}+\zeta_\mathbf{b})}\left(
      \Theta^{n}_{\mathbf{a}-1_q} - \Theta^{n+1}_{\mathbf{a}-1_q}
    \right), \label{eq:vrr} \\
    \Xi_{\mathbf{a}(\mathbf{b}+1_q)} &= \Xi_{(\mathbf{a}+1_q)\mathbf{b}} +
    (\mathbf{R}_\mathbf{a} - \mathbf{R}_\mathbf{b})_q\Xi_{\mathbf{ab}} \label{eq:hrr},
\end{align}
and
\begin{equation}
    \Theta^n_0 = F_n((\zeta_\mathbf{a}+\zeta_\mathbf{b})\vert \mathbf{R}_p-\mathbf{r}_i\vert^2), \qquad \Xi_\mathbf{a0} = \Theta^0_\mathbf{a},\qquad A_{\mathbf{ab} i} \equiv \Xi_{\mathbf{ab}}.
\end{equation}
$F_n$ is the Boys function defined in \cref{eq:boys}. Will refer to \cref{eq:vrr,eq:hrr} as the
vertical (VRR) and horizontal (HRR) OS recursions,
respectively, in the following. In the canonical two-step OS algorithm,
VRR intermediates are constructed and assembled into target integrals via
the HRR. For brevity, the following discussion considers the case that $\phi_\mathbf{a}$ and $\phi_\mathbf{b}$
are primitive GTO functions. This may be generalized to contracted functions by 
contracting the $\Theta$ intermediates with basis coefficients between the
VRR and HRR \cite{gill1989efficient}. There are two important aspects of these expressions for
the 3c-CP
\begin{enumerate}
    \item In contrast
to their application to the evaluation of ERIs\cite{gill1989efficient},
only the $\Theta^0$ intermediates are required for the assembly
of target integrals via the HRR. As such, for a particular 
$\mathbf{a}$ and $\mathbf{b}$, the VRR recursion
requires the evaluation of many intermediate quantities
which do not contribute to the HRR.

  \item Both the HRR and VRR are independent in the grid point, and are thus well suited to be parallelized in this dimension. On GPU architectures, each thread can independently perform the same VRR and HRR steps for different grid points (batch level parallelism) with minimal thread divergence. 
  However, adopting this simple approach can be resource (e.g. register and shared memory) intensive, which can hinder the number of thread blocks that can
  be executed concurrently on a particular SM.

\end{enumerate}
In the following,
we develop an efficient algorithm for the transversal of the VRR and HRR
recursions for the sn-K method to minimize GPU resource requirements.

\begin{figure}[t]
    \centering
    \includegraphics[width=0.8\textwidth]{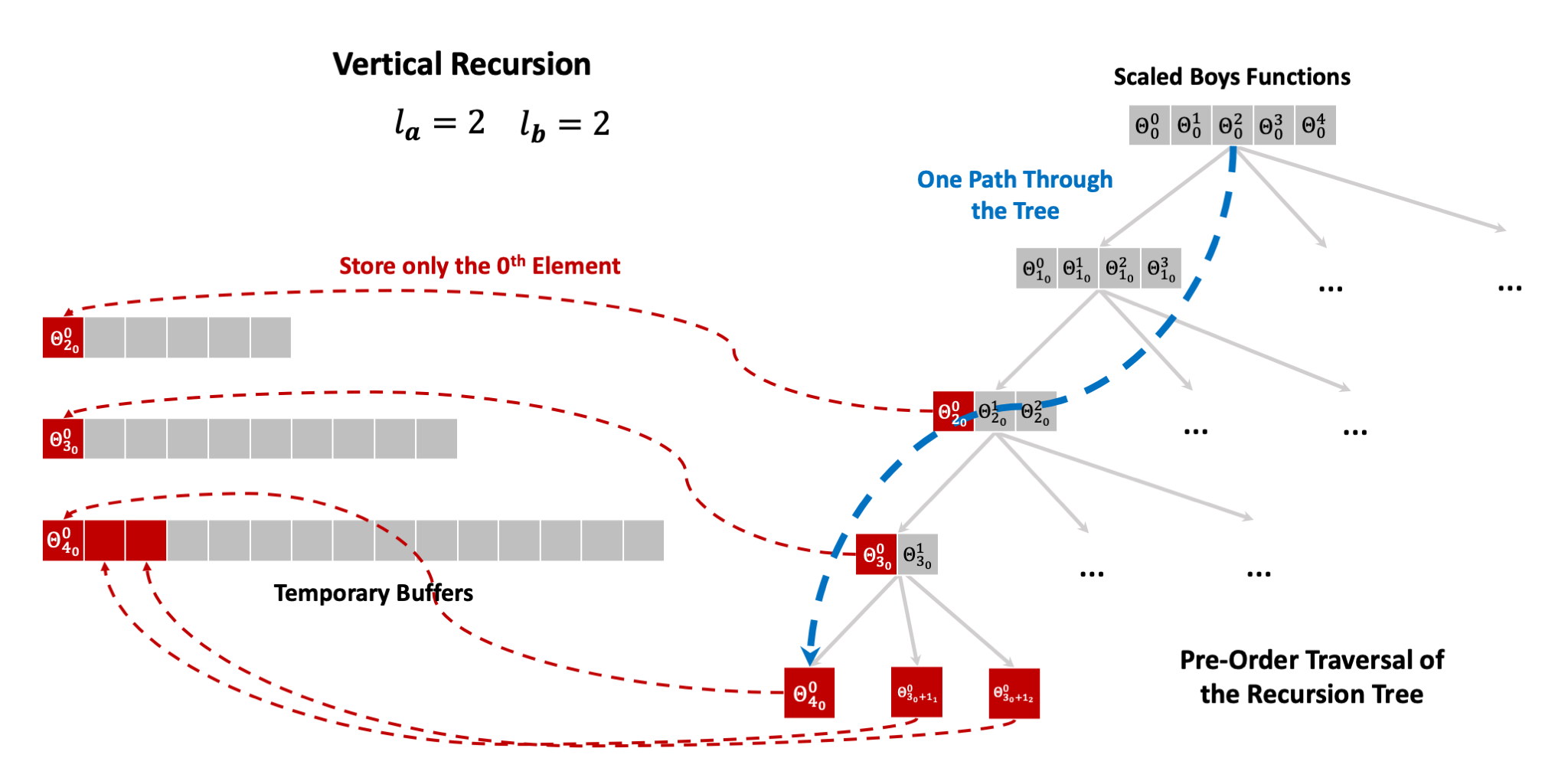}
    \caption{A pre-order tree transversal approach for the evaluation of the VRR expressed in \cref{eq:vrr}. The nodes and edges of this
    tree represent intermediate integrals and 
    and their VRR connections, respectively. Grey tiles represent temporary 
    terms which must be computed to form the 
    $\Theta$ values (red) required for the HRR.
    Only one path through the VRR tree transversal is shown for brevity.}
    \label{fig:vrr}
\end{figure}

{\bf{VRR Implementation.}} 
\Cref{fig:vrr} depicts our approach for implementing the VRR in the context of the sn-K method. First, we cast the recursion as a tree traversal, where the nodes of the tree represent the intermediate integrals and the edges between nodes represent the direct connections between the integrals via the VRR. The levels of the tree represent $\Theta^n$ intermediates with the same total angular momentum, $L$, with the root of the tree representing the state $L=0$. For a target $\mathbf{a}$ and $\mathbf{b}$, $L \in [0,l_\mathbf{a} + l_\mathbf{b} + 1]$, storage 
is kept to a minimum by performing a pre-order transversal of the VRR tree. This allows for the direct storage of $\Theta^0$ intermediates without having to form all elements of a tree level simultaneously. For example, via the pre-order transversal depicted in \cref{fig:vrr}, only $15$ $\Theta$ values need to be held simultaneously in $30$ single-precision GPU registers ($15$ x $2$ FP$64$) to evaluate every target integral for $l_\mathbf{a} = l_\mathbf{b} = 2$ as compared to $20$ $\Theta$ values for a breath first search traversal. Additionally, $30$ double precision intermediate values must be stored in shared memory, values that are needed for the HRR step. In general, our approach of the VRR step requires $\left(l_\mathbf{a} + l_\mathbf{b} + 1\right)\left(l_\mathbf{a} + l_\mathbf{b} + 2\right)$ single precision GPU registers, and $\sum^{l_\mathbf{b}}_{i = 1}\left(l_\mathbf{a} + i + 1\right)\left(l_\mathbf{a} + i + 2\right)$ single precision intermediate values in the GPU shared memory. The quantities must be multiplied with the total number of threads within an execution block to get the total resources needed. For small angular momenta (e.g., $l_\mathbf{a/b} < 2$), we do not utilize shared memory and keep all values in registers, includning the intermediate values. This has shown to provide the best performance at the expense of an increased register count.

\begin{figure}[t]
    \centering
    \includegraphics[width=0.8\textwidth]{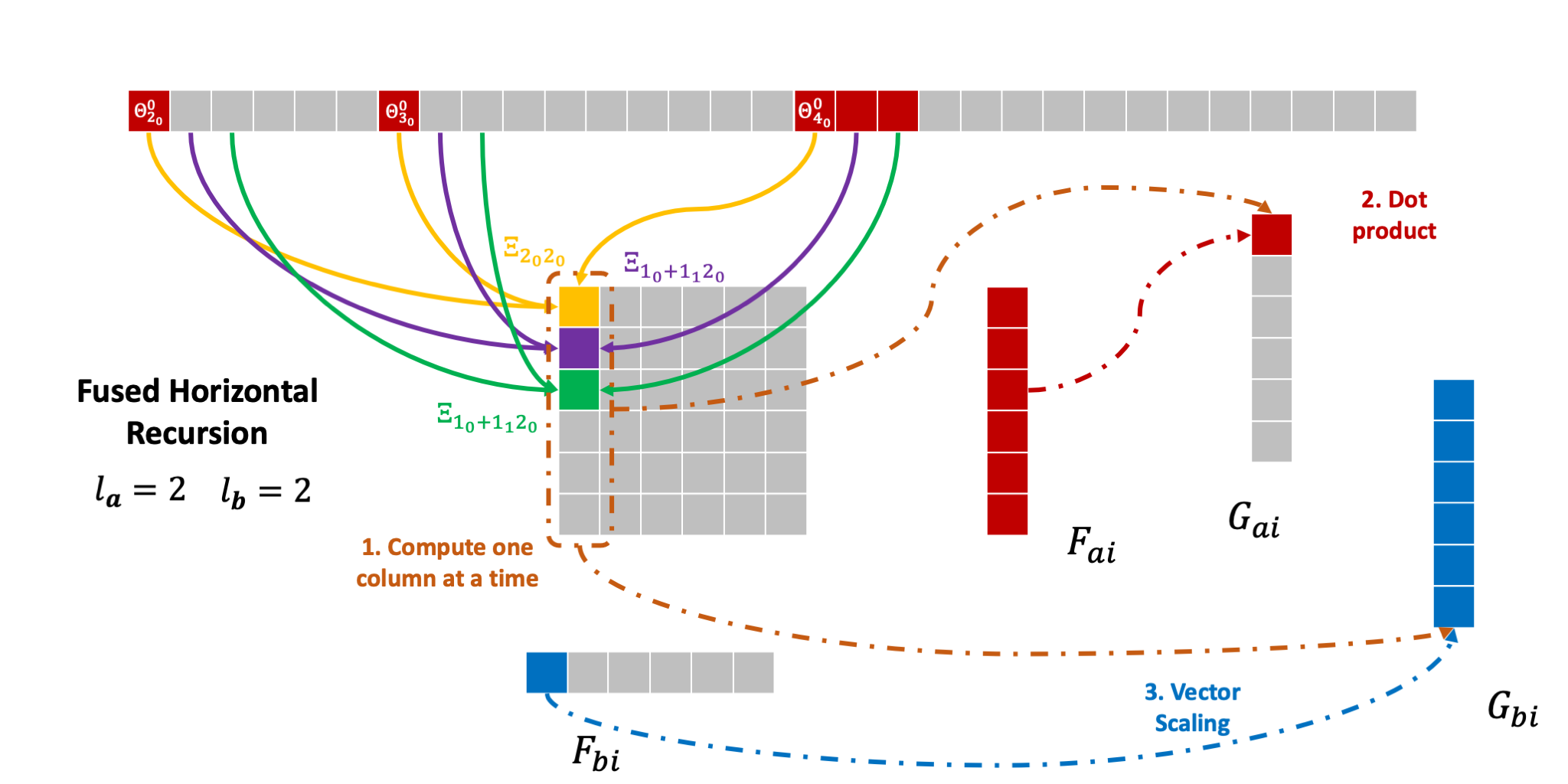}
    \caption{The construction of the $\Xi_{\mathbf{a}(\mathbf{b})}$ column by column and the subsequent merger with the contraction with the elements of $F_{\mathbf{a/b}i}$ to form $\mathbf{G}$. The full 3c-CP integral tensor is never fully materialized.}
    \label{fig:hrr}
\end{figure}

{\bf{Merged HRR Implementation.}} 
Given the required $\Theta^0$ intermediates from the VRR, Figure~\ref{fig:hrr} depicts our approach for merging the
HRR with the contraction with elements of $F_{\mathbf{a/b}i}$ to directly form its contributions to  $G_{\mathbf{a/b}i}$
in \cref{eq:gmat}. In this approach, the full 3c-CP integral tensor (which quadratically grows with angular momenta) need not be materialized in order to perform the contraction in \cref{eq:gmat}. For example, for $l_\mathbf{a}=l_\mathbf{b}=2$, the full 6x6 integral tensor occupies 72 single-precision (32 x 2 FP64) registers alone, which would constitute a significant bottleneck for resource utilization on modern GPU hardware, particularly when vectorizing over many grid points. In our approach, we only need to materialize a single column of the target integral at a time, which in the preceding example only requires $12$ registers (a $6$-fold reduction). Additionally, we need $36$ single precision registers to store the values from $F_{\mathbf{a/b}i}$ and the results for $G_{\mathbf{a/b}i}$. The values for $F_{\mathbf{a/b}i}$ can be pre-loaded to hide the latency of reading data from the GPU main memory. In general, for the HRR step our approach needs $2 *\left(l_\mathbf{a} + 1\right)\left(l_\mathbf{a} + 2\right) + \left(l_\mathbf{b} + 1\right)\left(l_\mathbf{a} + 2\right)$ single precision GPU registers. This values must be multiplied with the total number of threads within an thread block to obtain the final resource utilization.

For the GPU implementation, we have constructed device kernels for different angular momenta. More specifically we have constructed an in-house code generator that performs the pre-order traversal and fuses the horizontal recursion with the other contractions, and generates CUDA code. The current implementation exploits the independent execution across the grid points, each GPU thread executing the same VRR and HRR but on different grid points. For small angular momenta, we don't utilize shared memory and keep all intermediate values in registers. For larger values, we make use of shared memory, which is dynamically allocated given the angular momenta. Therefore the main strategy, given a shell pair of angular momenta $l_\mathbf{a}$ and $l_\mathbf{b}$, is to first group the grid points for that specific shell pair, launch a specialized kernel on the GPU and perform the batched VRR and HRR computations, and finally obtain the contracted $G_{\mathbf{a/b}i}$   stored in device memory.

\bibliography{grid_refs,sn_refs,refs,mb_refs,exa_refs,martinez_gpu_refs,ochsenfeld_gpu_refs,gamess_gpu_refs,quick_gpu_refs,my_refs.bib,vrgrefs.bib}

\end{document}